\pdfoutput=1
\documentclass[12pt]{article}

\usepackage{graphicx}

\input{epsf}

\def\rf#1{(\ref{eq:#1})}
\def\lab#1{\label{eq:#1}}
\def\nonu{\nonumber}
\def\br{\begin{eqnarray}}
\def\er{\end{eqnarray}}
\def\be{\begin{equation}}
\def\ee{\end{equation}}
\def\({\left(}
\def\){\right)}

\def\rlx{\relax\leavevmode}
\def\vp{\varphi}
\def\vpc{\varphi^*}

\def\ra{\rightarrow}
\def\ve{\varepsilon}

\newcommand{\sbr}[2]{\left\lbrack\,{#1}\, ,\,{#2}\,\right\rbrack}

\def\IZ{\rlx\hbox{\sf Z\kern-.4em Z}}
\def\IR{\rlx\hbox{\rm I\kern-.18em R}}
\def\IC{\rlx\hbox{\,$\inbar\kern-.3em{\rm C}$}}
\def\one{\hbox{{1}\kern-.25em\hbox{l}}}

\begin{document}

\begin{titlepage}
\vspace*{-1cm}

\vskip 2cm

\vspace{.2in}
\begin{center}
{\large\bf Quasi-Integrable Deformations of the Bullough-Dodd model}
\end{center}

\vspace{.5cm}

\begin{center}
Vinicius H. Aurichio~$^{\dagger}$  and  L. A. Ferreira~$^{\star}$

\vspace{.3 in}
\small

\par \vskip .2in \noindent
Instituto de F\'\i sica de S\~ao Carlos; IFSC/USP;\\
Universidade de S\~ao Paulo  \\ 
Caixa Postal 369, CEP 13560-970, S\~ao Carlos-SP, Brazil\\
$^{\dagger}$ e-mail: vinicius.aurichio@usp.br\\
$^{\star}$ e-mail: laf@ifsc.usp.br\\

\normalsize
\end{center}

%\vspace{.5in}

\begin{abstract}

\noindent It has been shown recently that  deformations of some integrable field theories in $(1+1)$-dimensions possess  an infinite number of charges that are asymptotically conserved in the scattering of soliton like solutions. Such charges are not conserved in time and they do vary considerably during the scattering process, however they all return in the remote future (after the scattering) to the values they had in the remote past (before the scattering). Such non-linear phenomenon was named {\em quasi-integrability}, and it seems to be related to special properties of the solutions under a space-time parity transformation. In this paper we investigate, analytically and numerically, such phenomenon in the context of deformations of the integrable Bullough-Dodd model. We find that a special class of two-soliton like solutions of such deformed theories do present an infinite number of asymptotically conserved charges.

\end{abstract} 
\end{titlepage}

\section{Introduction}
\label{sec:intro}
\setcounter{equation}{0}

The objective of the present paper is to investigate, analytically and numerically, the concept of quasi-integrability, first proposed in 
\cite{quasi1}, in the context of deformations of the exactly integrable Bullough-Dodd model \cite{bd}. 
The motivation of our study is to try to shed light on the mechanisms responsible for such interesting non-linear phenomenon which has a large potential for applications in many areas of physics,  mathematics and non-linear sciences in general.

As it is well known, solitons in $(1+1)$-dimensions are solutions of non-linear field equations which travel with constant speed without dispersion and dissipation, and when they scatter through each other they keep their forms, energies, etc, with the only effect being a shift in their positions relative to the ones  they would have if  the scattering have not occurred. Such extraordinary behavior is credited to the fact that the solitons appear in the so-called exactly integrable field theories in $(1+1)$-dimensions, that possess an infinite number of exactly conserved charges. Therefore, the only way for the scattering process to preserve the values of such an infinity of charges, is for the solitons to come out of it exactly as they have entered it. In addition, in most of such theories the strength of the interaction of the solitons is inversely related to the coupling constant, i.e. the solitons are weakly coupled in the strong regime and vice-versa. Such behavior and the large amount of symmetries (conserved charges) make the solitons the natural candidates for the normal modes of the theory in the strong coupling regime, opening the way for the development of many non-perturbative techniques in the study of non-linear phenomena. The drawback  of such approach is that exactly integrable soliton theories are rare, and few of them really describe phenomena in the real world. 

The observation put forward in \cite{quasi1} is that many theories which are not integrable present solutions that behave very similarly to solitons, i.e. such soliton like solution scatter through each other without distorting them very much. It was shown in \cite{quasi1} in the context of deformations of the sine-Gordon model, and then in other theories \cite{quasi2,quasi3,quasi4}, that such quasi-integrable theories possess and infinite number of charges that are asymptotically conserved. By that one means that during the scattering of two soliton like solutions such charges do vary in time (and quite a lot sometimes) but they all return in the remote future (after the scattering) to the values they had in the remote past (before the scattering). Since in a scattering process what matters are the asymptotic states, such theories are effectively integrable, and that is why they were named {\em quasi-integrable}. 

The mechanisms behind such non-linear phenomenon are not well understood yet. All the examples studied so far are deformations of  exactly integrable field theories. The zero-curvature condition or Lax-Zakharov-Shabat equation \cite{lax} of the integrable theory becomes anomalous when applied to the deformed theory, and so the Lax potentials fail to become flat connections when the equations of motion hold true. Despite those facts, techniques of integrable field theories can be adapted and applied to construct an infinite number of charges $Q^{(N)}$ which present an anomalous conservation law 
\be
\frac{d\,Q^{(N)}}{d\,t}=\beta^{(N)}
\lab{anomalousconserv}
\ee
The anomalies $\beta^{(N)}$ have some interesting properties. They vanish exactly when evaluated in one-soliton type solutions, and also vanish for two-soliton type solutions when the two solitons are well separated. The anomalies are only non-vanishing when the soliton like solutions are close together and interact with each other. For some special classes of soliton like solutions the anomalies $\beta^{(N)}$ have a further striking property. They have a mirror type symmetry in the sense that the charges have the same values when reflected around a particular value of time $t_{\Delta}$, which depends upon the parameters of the solution. In other words, one finds that $Q^{(N)}\({\tilde t}\)=Q^{(N)}\(-{\tilde t}\)$, where ${\tilde t}=t-t_{\Delta}$. So, the charges are not only asymptotically conserved, i.e. $Q^{(N)}\(\infty\)=Q^{(N)}\(-\infty\)$, but are symmetric with respect to a given value of the time. The only explanation found so far for such behavior of the charges, is that those special soliton like solutions transform in a special way under a space-time parity transformation, where the point in space-time around which space and time are reversed depend upon the parameters of the  solution under consideration. The proof of the connection between parity and mirror symmetry of the charges involves an interplay of the Lorentz transformations and internal transformations in the Kac-Moody algebra underlying the anomalous Lax equation.  We do not believe however that such parity property is  one of the causes of the quasi-integrability, but it seems to be present whenever such phenomenon occurs. 

In this paper we investigate the concept of quasi-integrability in the context of deformations of the Bullough-Dodd model \cite{bd} involving a  complex scalar field $\vp$ in $(1+1)$-dimensions with Lagrangian given by
\be
{\cal L} = \frac{1}{2}\, \partial_{\mu}\vp\, \partial^{\mu}\vp - V\(\vp\)
\lab{deformbdlag}
\ee 
and the potentials being 
\be
V\(\vp\)= e^{\vp}+\frac{1}{2+\ve}\, e^{-\(2+\ve\)\,\vp}
\lab{deformpot}
\ee
with $\ve$ being a  real deformation parameter, such that the Bullough-Dodd model is recovered in the case $\ve=0$. Because of some particularities of the vacuum solutions of such models, as explained in section \ref{sec:intro}, the physically interesting deformed theories exist only when the  parameter $\ve$ is restricted to rational values. 

We construct the anomalous zero-curvature condition for the theories \rf{deformbdlag} with the Lax potentials taking values on the twisted $sl(2)$ Kac-Moody algebra $A_2^{(2)}$. The charges 
$Q^{(N)}$ satisfying \rf{anomalousconserv} are obtained by  the so-called abelianization procedure  \cite{olive1,olive2,aratyn,massformula,drinfeld} where the Lax potentials are gauge transformed into an infinite abelian sub-algebra of $A_2^{(2)}$. In  fact, due to the anomaly of the zero-curvature only one component of the Lax potentials can be rotated into such sub-algebra, leading therefore to the anomalous conservation  \rf{anomalousconserv}. The Lax potentials do not transform as vectors under the $(1+1)$-dimensional Lorentz transformations. However, the grading operator of the Kac-Moody algebra $A_2^{(2)}$ generates a one-dimensional subgroup isomorphic to the Lorentz group, and we show that the Lax potentials are vectors under the combined action of those two groups. That fact allows us to show that the anomalies in \rf{anomalousconserv} do vanish when evaluated on the one-soliton solutions of the theories \rf{deformbdlag}. In addition, we show that some special two-soliton solutions lead to the existence of a  space-time parity transformation $P$ such that the complex scalar field $\vp$, when evaluated on them, transforms as $P\(\vp\)=\vp^*$. For such two-soliton solutions we show that the real part of the charges $Q^{(N)}$ satisfy a mirror symmetry, as described above, and are therefore asymptotically conserved. The imaginary part of the charges however, are not asymptotically conserved.

We also implement a perturbative method to construct solutions of the deformed theory  
\rf{deformbdlag} as a power series  in the deformation parameter $\ve$, as $\vp=\vp_0+\ve\,\vp_1+\ve^2\,\vp_2+\ldots$, such that $\vp_0$ is an exact solution of the integrable Bullough-Dodd model. We then split the fields into their real and imaginary parts and then into their even and odd parts under the parity transformation $P$, i.e. $\vp^{R/I,\pm}_n=\frac{1}{2}\(1\pm P\)\vp^{R/I}_n$, with $\vp_n=\vp^R_n+i\,\vp^I_n$. By starting with an exact  solution $\vp_0$ of the Bullough-Dodd model that satisfies $P\(\vp_0\)=\vp_0^*$, or $P\(\vp_0^R\)=\vp_0^R$ and  $P\(\vp_0^I\)=-\vp_0^I$, we show that the pair of fields $\(\vp_1^{R,+}\, , \, \vp_1^{I,-}\)$ satisfy a pair of linear non-homogeneous equations, and the pair of fields $\(\vp_1^{R,-}\, , \, \vp_1^{I,+}\)$ satisfy a pair of linear homogeneous equations. Therefore, it is always possible to choose solutions where $\(\vp_1^{R,-}\, , \, \vp_1^{I,+}\)=0$, and so the first order field satisfies, $P\(\vp_1\)=\vp_1^*$. Once that is chosen, one can show that the same structure repeats at the second order in the expansion and one can choose solutions such that $P\(\vp_2\)=\vp_2^*$. By repeating such procedure order by order we show that the theories  \rf{deformbdlag} always contain solutions with the property $P\(\vp\)=\vp^*$, and so the charges evaluated on them present the mirror symmetry described above. So, the dynamics of the deformed theories \rf{deformbdlag} favors the ``good'' solutions, in the sense that it is not possible to have solutions satisfying the pure opposite behavior under the parity, i.e. $P\(\vp\)=-\vp^*$. That is an interesting interplay between the dynamics and the parity that deserves further study. For instance, the production of ``bad'' modes could be energetically disfavored and emission of radiation could be suppressed. 
  
We also perform numerical simulations, based on the fourth order Runge-Kutta method, to study the scattering of two soliton like solutions of \rf{deformbdlag}. We performed simulations for various rational values of the deformation parameter $\ve$, and found that in all cases the predictions of the analytical calculations were confirmed, i.e. the real part of charges do satisfy the mirror symmetry when we use as a seed for the simulations solutions of the Bullough-Dodd model that have the right parity property, i.e. $P\(\vp\)=\vp^*$. So, the evolution of the fields under the deformed equations of motion seem not to destroy the parity property of the initial configuration, again indicating that the dynamics seem to favor the ``good'' modes as observed in the analytical perturbative expansion mentioned above. The mirror symmetry of the charges were checked in the simulations by evaluating the first non-trivial anomaly $\beta^{(5)}$ (see \rf{anomalousconserv}) as well as its integrated version $\gamma^{(5)}= \int_{-\infty}^t dt^{\prime}\,   \beta^{(5)}=Q^{(5)}\(t\)-Q^{(5)}\(-\infty\)$. All simulations show that the real part of $\gamma^{(5)}$ is symmetric under reflection around a given value of time close to $t=0$, and so leading to the mirror symmetry for the real part of the charge $Q^{(5)}$, and then for its asymptotic conservation. The imaginary part of $\gamma^{(5)}$ is not symmetric under reflection and does not lead to the asymptotic conservation of the imaginary part of the charge $Q^{(5)}$. 

The paper is organized as follows: section \ref{sec:bdmodels} discusses the properties of the vacuum solutions of the theories \rf{deformbdlag} and their implications on the possible physically interesting deformations of the Bullough-Dodd model. In section \ref{sec:charges} we present the construction of the quasi-conserved charges using techniques of integrable field theories based on the anomalous zero-curvature or Lax-Zakharov-Shabat equation. The interplay between the Lorentz and parity transformations leading to the mirror symmetry of the charges is dicussed in section \ref{sec:lorentz}, and section \ref{sec:parityversusanomaly} implements the perturbative method to construct solutions of \rf{deformbdlag} as power series in the deformation parameter $\ve$ and discusses the connection between dynamics and parity. The Hirota's one-soliton and two-soliton exact solutions of the integrable Bullough-Dodd model are given in section  \ref{sec:hirota}. The numerical simulations are presented in section \ref{sec:simulations}, and our conclusions are given in section \ref{sec:conclusions}. The appendix \ref{app:kacmoody} gives some basic results about the twisted $sl(2)$ Kac-Moody algebra $A_2^{(2)}$ used in the text.

\section{The deformed Bullough-Dodd models}
\label{sec:bdmodels}
\setcounter{equation}{0}

We shall consider models of a complex scalar field $\vp$ in $(1+1)$-dimensions with Lagrangian given by \rf{deformbdlag}  and the potentials being given by \rf{deformpot}. The Euler-Lagrange equation following from \rf{deformbdlag} is
\be
\partial^2_t \vp - \partial^2_x \vp+ e^{\vp}- e^{-\(2+\ve\)\,\vp}=0
\lab{deformeqom}
\ee
where we have taken the speed of light to be unity. The  real and imaginary parts of the equation 
\rf{deformeqom} are given by
\br
\partial^2_t \vp_R - \partial^2_x \vp_R +e^{\vp_R}\, \cos\(\vp_I\) - 
e^{-\(2+\ve\)\,\vp_R}\, \cos\(\(2+\ve\)\,\vp_I\)&=&0
\nonumber\\
\partial^2_t \vp_I - \partial^2_x \vp_I +e^{\vp_R}\, \sin\(\vp_I\) + 
e^{-\(2+\ve\)\,\vp_R}\, \sin\(\(2+\ve\)\,\vp_I\)&=&0
\lab{deformeqom2}
\er
where we have denoted  $\vp = \vp_R+i\,\vp_I$. The Hamiltonian associated to \rf{deformbdlag} is conserved and complex, and denoting the Hamiltonian density as ${\cal H}={\cal H}_R+i\,{\cal H}_I$,  we get
\br
{\cal H}_R&=& \frac{1}{2}\left[\(\partial_t\vp_R\)^2+ \(\partial_x \vp_R\)^2-\(\partial_t\vp_I\)^2-\(\partial_x \vp_I\)^2\right] + V_R
\nonumber\\
{\cal H}_I&=&\partial_t\vp_R\,\partial_t\vp_I +\partial_x \vp_R\, \partial_x \vp_I + V_I
\lab{hamdensity}
\er
with $V=V_R+i\, V_I$, and 
\br
 V_R&=& e^{\vp_R}\, \cos \(\vp_I\)+\frac{1}{2+\ve}\,e^{-\(2+\ve\)\,\vp_R}\,\cos\(\(2+\ve\)\,\vp_I\)
 \nonumber\\
 V_I&=& e^{\vp_R}\, \sin \(\vp_I\) -\frac{1}{2+\ve}\,e^{-\(2+\ve\)\,\vp_R}\,\sin\(\(2+\ve\)\,\vp_I\)
 \lab{deformpot2}
 \er
Note that the densities \rf{hamdensity} are not positive definite, and so we can not really define vacuum  configurations as those having minimum energies. However, in order for the space integrals of the densities \rf{hamdensity} to be conserved in time, one needs the momenta to vanish at spatial infinity, and so one needs the fields to be constants there. Therefore, from the equations of motion 
\rf{deformeqom2} one gets that such constant configurations are extrema of the potentials, i.e.
\br
\frac{\partial V_R}{\partial\vp_R}=\frac{\partial V_I}{\partial\vp_I}&=&
e^{\vp_R}\, \cos\(\vp_I\) - e^{-\(2+\ve\)\,\vp_R}\, \cos\(\(2+\ve\)\,\vp_I\)=0
\nonumber\\
\frac{\partial V_I}{\partial\vp_R}=-\frac{\partial V_R}{\partial\vp_I}&=&
e^{\vp_R}\, \sin\(\vp_I\) + e^{-\(2+\ve\)\,\vp_R}\, \sin\(\(2+\ve\)\,\vp_I\)=0
\lab{extrema}
\er
which implies
\be
e^{\(3+\ve\)\,\vp_R}\, \cos\(\vp_I\) =  \cos\(\(2+\ve\)\,\vp_I\) \qquad\qquad
e^{\(3+\ve\)\,\vp_R}\, \sin\(\vp_I\) =-  \sin\(\(2+\ve\)\,\vp_I\)
\lab{extrema2}
\ee
Squaring both equations and adding them up one concludes  that $\vp_R=0$. Using that fact and  manipulating \rf{extrema2}, by multiplying and adding them up,  one concludes that $\sin\(\(3+\ve\)\,\vp_I\)=0$ and $\cos\(\(3+\ve\)\,\vp_I\)=1$. Therefore, the extrema of the potentials are
\be
\(\vp_R\, , \, \vp_I\)=\(0\, , \, \frac{2\,\pi\, n}{3+\ve}\) \qquad\qquad \qquad \qquad \mbox{\rm with  $n$  integer}
\lab{vacua}
\ee
Note that such extrema are not maxima or minima of the potentials. They all correspond to saddle points. 

We now come to a very interesting property of such models. For static configurations we have that the quantities 
\br
{\cal E}_R= -\frac{1}{2}\left[ \(\partial_x \vp_R\)^2-\(\partial_x \vp_I\)^2\right] + V_R
\; ; \qquad\qquad 
{\cal E}_I=-\partial_x \vp_R\, \partial_x \vp_I + V_I
\lab{hamdensitystatic}
\er
are constant in $x$ as a consequence of the static equations of motion \rf{deformeqom2}, i.e. 
$\frac{d\,{\cal E}_R}{d\, x}=\frac{d\,{\cal E}_I}{d\, x}=0$. Indeed, those quantities correspond to the static Hamiltonian densities \rf{hamdensity} with  the space coordinate $x$ replaced by an imaginary time $i\, x$. So, the quantities \rf{hamdensitystatic} are conserved  in the time $i\,x$ for a mechanical problem of a particle moving on a two dimensional space with coordinates $\vp_R$ and $\vp_I$. Therefore, for a given static solution of the model \rf{deformbdlag} that goes, at spatial infinity,  to vacua configurations \rf{vacua}, one concludes that the potentials $V_R$ and $V_I$ (and so the complex potential $V$) are bound  to have the same values at both ends of spatial infinity.  If one denotes by $n_{+}$ and $n_{-}$ the integers labeling the vacua \rf{vacua} at $x\rightarrow \infty$ and $x\rightarrow -\infty$, respectively, then from \rf{deformpot} one gets that 
\be
e^{\frac{i\,2\,\pi\, \(n_{+}-n_{-}\)}{\(3+\ve\)}}=1
\lab{nicecond}
\ee
For the Bullough-Dodd model where $\ve=0$, one gets that $ n_{+}-n_{-}$ has to be a multiple of $3$. Indeed, the static one-soliton solutions of that model do not ``tunnel'' between consecutive vacua as $x$ varies from $-\infty$ to $+\infty$, like in the sine-Gordon model, but jumps $2$ vacua and ends in the third one. 

For the deformed Bullough-Dodd models where $\ve\neq 0$, the only way of satisfying the condition 
\rf{nicecond} for any real value of $\ve$ is to have $ n_{+}=n_{-}$, i.e. for a static solution the vacua are the same at both ends of spatial infinity. If one wants non-trivial one soliton solution with non vanishing topological charge, then the deformation parameter $\ve$ has to be taken to be a  rational number. That is a quite striking and interesting restriction on the ways the Bullough-Dodd model can be deformed. If one takes 
\be
2+\ve = \frac{p}{q} \qquad\qquad\qquad \mbox{\rm with $p$ and $q$ integers}
\ee
then the equations of motion \rf{deformeqom}  takes the form
\be
\partial^2_{\hat t} \phi - \partial^2_{\hat x} \phi+ e^{q\,\phi}- e^{-p\,\phi}=0
\lab{deformeqomrational}
\ee
 where we have redefined the fields as  $\vp \equiv q\, \phi$, and  the space-time coordinates as $x \equiv  \sqrt{q}\, {\hat x}$, and  $t\equiv \sqrt{q}\, {\hat t}$.

\section{The quasi-conserved charges}
\label{sec:charges}
\setcounter{equation}{0}

We now use techniques of exactly integrable field theories in $(1+1)$ dimensions to construct quasi-conserved charges for the non-integrable theories we are considering.  We introduce the connection (Lax potentials) 
\be
A_{+}= \(-2\, V+m\)\, b_1 + \frac{\partial \,V}{\partial\,\vp}\, F_1\; ; \qquad\qquad\qquad 
A_{-}= b_{-1}-\partial_{-}\,\vp\, F_0
\lab{deformedpot}
\ee
where we have used light-cone coordinates as $x_{\pm} \equiv \frac{ t\pm x}{2}$, and so 
$ \partial_{\pm} = \partial_t \pm \partial_x$.  In addition,  the operators $b_n$ and $F_n$ are generators of the twisted loop (Kac-Moody) algebra $A_2^{(2)}$  defined in appendix 
\ref{app:kacmoody}. Even though we will be interested in potentials of the form \rf{deformpot}, in the connection \rf{deformedpot} we shall assume that $V$ is a generic potential that depends upon the scalar field $\vp$ but not on its complex conjugate $\vp^*$. In addition, we shall assume that the potential does not involve complex parameters in its definition, such that its complex conjugate can be obtained just by the replacement $\vp \ra \vpc$, i.e. 
\be
V^*=V\(\vp\ra \vpc\)
\lab{conjugatepot}
\ee
The reasons for assuming that property will become clear when we discuss below the anomalies in the conservation of the charges.  

One can check that the curvature of the connection \rf{deformedpot} is given by 
\br
\partial_{+}A_{-}-\partial_{-}A_{+}+\sbr{A_{+}}{A_{-}}=
-\(\partial_{+}\partial_{-}\vp+\frac{\partial \,V}{\partial\,\vp}\)\,F_0
 -  \partial_{-}\vp\,X\, F_1
 \lab{deformedcurvature}
\er
with 
\be
X=\frac{\partial^2 \,V}{\partial\,\vp^2}+\frac{\partial \,V}{\partial\,\vp}-2\,V+m
\lab{xdef}
\ee
The coefficient of $F_0$ in \rf{deformedcurvature} corresponds to the equation of motion of the theory
\be
\partial_{+}\partial_{-}\vp+\frac{\partial \,V}{\partial\,\vp}=0
\lab{eqofmotgenv}
\ee
and, when it holds true the vanishing of the curvature depends upon the vanishing of the anomalous term $X$. Note that by shifting the potential as $V\rightarrow V+\frac{m}{2}$, one observes that $X$ vanishes only when $V$ is a linear combination of the exponential terms $e^{-2\,\vp}$ and $e^{\vp}$. So, the curvature \rf{deformedcurvature} vanishes for the Bullough-Dodd potential, corresponding to 
\rf{deformpot} for $\ve=0$, or then for the potential of the Liouville model,  $V\sim e^{\vp}$. The fact that the Liouville model admits a zero curvature representation in terms of the twisted Kac-Moody algebra $A_2^{(2)}$ is perhaps not know in the literature.

In order to construct the charges we use the so-called abelianization procedure \cite{olive1,olive2,aratyn,massformula} or Drinfeld-Sokolov reduction \cite{drinfeld}. We perform a gauge transformation of the deformed connection \rf{deformedpot} as
\be
A_{\mu}\rightarrow a_{\mu}= g\, A_{\mu}\,g^{-1}-\partial_{\mu} g\, g^{-1}
\lab{rotatedamu}
\ee
with $g$ being an exponentiation of the positive grade generators $F_n$, introduced in the appendix 
\ref{app:kacmoody}, 
\be
g= e^{\sum_{n=1}^{\infty} {\cal F}_{n}} \qquad\qquad {\cal F}_{n} \equiv \zeta_{n}\, F_n
\ee
Splitting things according to the grading operator \rf{gradingop}, we have that the $a_{-}$ component of the transformed potential \rf{rotatedamu}, becomes 
\be
a_{-}=\sum_{n=-1}^{\infty} a_{-}^{(n)}\; ; \qquad\qquad\qquad \sbr{d}{a_{-}^{(n)}}=n\,a_{-}^{(n)}
\ee
The components of it are 
\br
a_{-}^{(-1)} &=& b_{-1}\nonumber\\
a_{-}^{(0)} &=& -\sbr{b_{-1}}{{\cal F}_{1}}- \partial_{-}\varphi\, F_0
\lab{splitrotatedpot}\\
a_{-}^{(1)} &=& -\sbr{b_{-1}}{{\cal F}_{2}}+\frac{1}{2}\,\sbr{\sbr{b_{-1}}{{\cal F}_{1}}}{{\cal F}_{1}} -  \partial_{-}\varphi\, \sbr{{\cal F}_1}{F_0}-\partial_{-}{\cal F}_1
\nonumber\\
&\vdots&\nonumber\\
a_{-}^{(n)} &=& -\sbr{b_{-1}}{{\cal F}_{n+1}}+ \ldots
\nonumber
\er
The crucial algebraic property used in the following calculation is the fact that $b_{-1}$ is a semi-simple element of the algebra ${\cal G}=A_2^{(2)}$, in the sense that ${\cal G}$ is split into the kernel and image of  its adjoint action and they have an empty intersection, i.e. 
\be
{\cal G}= {\rm Ker}+ {\rm Im} \qquad \qquad \sbr{b_{-1}}{{\rm Ker}}=0\qquad\qquad 
{\rm Im}=\sbr{b_{-1}}{{\cal G}}
\lab{kernelimage}
\ee
From the appendix \ref{app:kacmoody} one observes that the elements of ${\rm Ker}$ are $b_{6n\pm  1}$, and the elements of ${\rm Im}$ are $F_n$. Since ${\cal F}_{n+1}$ is in the image of the adjoint action of $b_{-1}$, none of its components commute with $b_{-1}$ . Therefore one can recursively chooses the parameters $\zeta_{n+1}$ inside the ${\cal F}_{n+1}$ to kill the component of 
$a_{-}^{(n)}$ in the image, i.e. in the direction of $F_{n}$.  After that procedure is done the connection $a_{-}$ takes the form
\be
a_{-}= b_{-1}+ a_{-}^{b,(1)}\, b_1+ a_{-}^{b,(5)}\, b_5 + \ldots
\ee
The first two non-trivial components  are
\be
a_{-}^{b,(1)}=\(\partial_{-}\varphi\)^2
\ee
and
\br
a_{-}^{b,(5)}&=& \frac{1}{3}  \(\partial_{-}\varphi\)^6-\frac{23}{36}  
   \(\partial_{-}^2\varphi\) \(\partial_{-}\varphi\)^4-\frac{10}{3} 
   \(\partial_{-}^3\varphi\) \(\partial_{-}\varphi\)^3
   \nonumber\\
   &-&5  
   \(\partial_{-}^2\varphi\)^2 \(\partial_{-}\varphi\)^2+\frac{2}{3} 
   \(\partial_{-}^4\varphi\) \(\partial_{-}\varphi\)^2+\frac{14}{3}
    \(\partial_{-}^2\varphi\) \(\partial_{-}^3\varphi\) 
   \(\partial_{-}\varphi\)
   \nonumber\\
   &+& \partial_{-}^5\varphi  
   \(\partial_{-}\varphi\)
\er
Once the parameters $\zeta_n$ are chosen to rotate $a_{-}$ into the abelian subalgebra (kernel) generated by $b_{6n\pm 1}$, there is nothing we can do about the $a_{+}$ component of the transformed connection \rf{rotatedamu}. We only know it will have positive grade components only, since $g$ is generated by the positive grade $F_n$'s. Therefore, it is of the form 
\be
a_{+}= \sum_{N=1}^{\infty} a_{+}^{b,(N)}\, b_{N} +\sum_{n=1}^{\infty} a_{+}^{F,(n)}\, F_n\; ;
\qquad\qquad\qquad
N=6\,l\pm 1\; ; \qquad l=0, 1, 2, \ldots
\lab{Ndefinaplus}
\ee
In the case of integrable field theories where the quantity $X$, defined in \rf{xdef}, vanishes, it is possible to show that by using the equations of motion the image component of $a_{+}$ vanishes. In our case, the use of the equations of motion \rf{eqofmotgenv} can show that all the quantities $a_{+}^{F,(n)}$ are linear in $X$, given in \rf{xdef}. 

When the equations of motion  \rf{eqofmotgenv} hold true the transformed curvature reads (see \rf{deformedcurvature})
\be
\partial_{+}a_{-}-\partial_{-}a_{+}+\sbr{a_{+}}{a_{-}}= - \partial_{-}\varphi\, X\, g\, F_1\, g^{-1}
\lab{rotatedcurvature}
\ee
 We now write
\be
g\, F_1\, g^{-1}=\sum_{N=1}^{\infty} \alpha^{(N)}\, b_{N} +\sum_{n=1}^{\infty} \beta^{(n)}\, F_n
\lab{rotatef1}
\ee
and the first two components of the first term on the r.h.s. of  \rf{rotatef1} are 
\br
\alpha^{(1)}&=&0
\lab{anomalyfinal}\\
\alpha^{(5)}&=&4  \(\partial_{-}\varphi\)^2 \(\partial_{-}^2\varphi\)-4 
   \(\partial_{-}^2\varphi\)^2-2  \(\partial_{-}\varphi\) 
   \(\partial_{-}^3\varphi\)-2  \(\partial_{-}^4\varphi\)
   \nonumber
\er
Since $a_{-}$ does not have components in the direction of the $F_n$'s, it follows that the commutator in \rf{rotatedcurvature} does not produce terms in the directions of the $b_N$'s. Therefore, one has that
\be
\partial_{+}a_{-}^{b,(N)}-\partial_{-}a_{+}^{b,(N)}= - \partial_{-}\varphi\, X\,\alpha^{(N)}
\ee
which in terms of the space-time coordinates $x$ and $t$ becomes 
\be
\partial_{t}a_{x}^{b,(N)}-\partial_{x}a_{t}^{b,(N)}= \frac{1}{2}\, \partial_{-}\varphi\, X\,\alpha^{(N)}
\ee
Defining the charges as
\be
Q^{(N)}\equiv \int_{-\infty}^{\infty} dx\, a_{x}^{b,(N)} 
\lab{chargesdef}
\ee
we get 
\be
\frac{d\,Q^{(N)}}{d\,t}= \beta^{(N)} \qquad\qquad\qquad 
\beta^{(N)}\equiv \frac{1}{2}\, \int_{-\infty}^{\infty} dx\,\partial_{-}\varphi\, X\,\alpha^{(N)}
\lab{quasiconserved}
\ee
where we have assumed boundary conditions such that $a_{t}^{b,(N)}\(x\ra \infty\)-a_{t}^{b,(N)}\(x\ra -\infty\)=0$. We call $\beta^{(N)}$ the anomaly of the charge $Q^{(N)}$. An useful quantity in our numerical simulations is what we call the integrated anomaly $\gamma^{(N)}$ defined as 
\be
\gamma^{(N)}\equiv Q^{(N)}\(t\)-Q^{(N)}\(-\infty\)= \frac{1}{2}\, \int_{-\infty}^{t} dt^{\prime}\,\int_{-\infty}^{\infty} dx\,\partial_{-}\varphi\, X\,\alpha^{(N)}
\lab{intanomaly}
\ee
From \rf{anomalyfinal} one notes that the anomaly $\beta^{(1)}$ vanishes and so the charge $Q^{(1)}$ is conserved. It corresponds in fact to a linear combination of the energy and momentum. 
The first non trivial anomaly corresponds to $N=5$ and the expression for $\alpha^{(5)}$ and $X$ are given in  \rf{anomalyfinal} and \rf{xdef} respectively. 

\section{The role of Lorentz and parity transformations}
\label{sec:lorentz}
\setcounter{equation}{0}

Consider the $(1+1)$-dimensional Lorentz  transformation
\be
\Lambda: \qquad x_{\pm}\ra e^{\mp \alpha}\,x_{\pm} \qquad\qquad\quad {\rm or} \qquad\qquad 
x\ra \frac{x-v\,t}{\sqrt{1-v^2}}\; ; \qquad t\ra \frac{t-v\,x}{\sqrt{1-v^2}}
\ee
where $\alpha$ is the rapidity and $v$ the velocity, i.e. $v=\tanh \alpha$. The Lax potentials 
\rf{deformedpot} do not transform as vectors under such Lorentz boost. However, consider the  automorphism of  the loop algebra $A_2^{(2)}$
\be
\Sigma\(T\) = e^{\alpha\, d}\,T\, e^{-\alpha\,d}
\ee
where $d$ is the grading operator defined in \rf{gradingop}. It then follows that the Lax operators 
\rf{deformedpot} transform, under the composed transformation, as vectors, i.e. 
\be
\Omega\(A_{\pm}\)=  e^{\pm \alpha}\,A_{\pm}\qquad\qquad \qquad {\rm with}\qquad\qquad 
\Omega\equiv\Lambda\,\Sigma
\lab{omegadefapm}
\ee
Therefore, the curvature \rf{deformedcurvature} is invariant under such combined transformation, and so is the anomalous term $\partial_{-}\vp\,X\, F_1$. In order to see how the anomalies of the charges \rf{chargesdef} transform under $\Omega$, we have to inspect the properties of  the quantities $\alpha^{(N)}$, introduced in \rf{rotatef1}. 

Note that the term $\partial_{-}\varphi\, F_0$, appearing on the r.h.s. of the second equation of \rf{splitrotatedpot}, transforms  under $\Omega$ as
\be
\Omega\(\partial_{-}\varphi\, F_0\) = e^{\alpha}\, \partial_{-}\varphi\, F_0
\ee
As explained below \rf{splitrotatedpot}, we choose the parameter $\zeta_1$ in ${\cal F}_{1}$ such that the term $\sbr{b_{-1}}{{\cal F}_{1}}$ cancels the term $\partial_{-}\varphi\, F_0$, and so such two terms have to  transform under the same  rule under $\Omega$. Since $\Omega\(b_{-1}\)=e^{\alpha}\,b_{-1}$, it follows that 
\be
\Omega\( {\cal F}_{1}\)={\cal F}_{1}
\lab{f1omega}
\ee
Indeed, one finds that the cancelation implies that one should choose $\zeta_1=-\partial_{-}\varphi$, and so $\Omega\(F_1\)=e^{\alpha}\,F_1$, and $\Omega\(\zeta_1\)=e^{-\alpha}\,\zeta_1$. Now, using 
\rf{f1omega} one observes that the last three terms on the r.h.s. of the third equation in  
\rf{splitrotatedpot} gets multiplied by $e^{-\alpha}$ under the action of $\Omega$. Therefore, in order to cancel the $F_1$ component on the r.h.s. of that equation one needs $\Omega\(\sbr{b_{-1}}{{\cal F}_{2}}\)=e^{-\alpha}\, \sbr{b_{-1}}{{\cal F}_{2}}$, and so one has to have $\Omega\( {\cal F}_{2}\)={\cal F}_{2}$. Continuing such reasoning order by order, one concludes that all ${\cal F}_{n}$'s have to be invariant under $\Omega$, and so the group element performing the gauge transformation 
\rf{rotatedamu} is also invariant, i.e. 
\be
\Omega\( g\)=g
\lab{gomega}
\ee
Therefore, similarly to $A_{\mu}$, the transformed connection $a_{\mu}$ also behaves as vector under $\Omega$, i.e.
\be
\Omega\(a_{\pm}\)= e^{\pm \alpha}\, a_{\pm}
\ee
According to the way the ${\cal F}_n$'s are chosen in \rf{splitrotatedpot} to cancel the components of $a_{-}$ in the direction of the $F_n$'s, it follows that the parameters $\zeta_n$'s are functions of the $x_{-}$-derivatives of the scalar field $\vp$. But from \rf{gomega} one concludes that
\be
\Omega\(\zeta_n\)=\Lambda\(\zeta_n\)=e^{-n\, \alpha} \,\zeta_n
\lab{zetatransform}
\ee
Therefore, each term in $\zeta_n$ must contain $n$ $x_{-}$-derivatives of the field $\vp$. 
From \rf{gomega} one observes that $\Omega\(g\, F_1\, g^{-1}\)= e^{ \alpha}\,g\, F_1\, g^{-1}$, and so every term on the r.h.s. of \rf{rotatef1} have to get mulplied by $e^{ \alpha}$ under the action of $\Omega$. Therefore, since $\Omega\(b_N\)=  e^{N\, \alpha}\, b_N$, it follows that
\be
\Omega\(\alpha^{(N)}\)=\Lambda\(\alpha^{(N)}\)=e^{\(-N+1\)\, \alpha} \,\alpha^{(N)}
\lab{alphatransform}
\ee
From the definition of $\alpha^{(N)}$ in \rf{rotatef1}, it is clear that it is a function of the $\zeta_n$'s, and so a function of the $x_{-}$-derivatives of the field $\vp$. Therefore, from \rf{alphatransform} one concludes that each term in $\alpha^{(N)}$ must contain $(N-1)$ $x_{-}$-derivatives of $\vp$. Indeed, from \rf{anomalyfinal} one observes that $\alpha^{(5)}$  contains four $x_{-}$-derivatives of $\vp$.

We are now in a position to draw some conclusions about the anomalies of the charges $Q^{(N)}$ defined in  \rf{chargesdef}. Consider a solution of the equations of motion \rf{eqofmotgenv} which is a traveling wave, i.e. $\vp=\vp\(x-v\,t\)$. One can then  make a Lorentz transformation and go to the rest frame of such solution where it becomes static, i.e. $x$-dependent only. Clearly the charges  $Q^{(N)}$ evaluated on such solution must be time independent, and so its anomaly $\beta^{(N)}$, defined in 
\rf{quasiconserved}, must vanish.  But from \rf{quasiconserved} and \rf{alphatransform} it follows that 
\be
\Omega\(\beta^{(N)}\,dt\)=\Lambda\(\beta^{(N)}\,dt\)= e^{-N\, \alpha} \,\beta^{(N)}\,dt
\lab{tensoranomaly}
\ee
Therefore, $\beta^{(N)}\,dt$, and so $dQ^{(N)}$, is a tensor under the $(1+1)$-dimensional Lorentz group. Consequently, if $dQ^{(N)}$ vanishes on the rest frame of the solution, it should vanish in all Lorentz frames. One then concludes that the charges $Q^{(N)}$ are exactly conserved for traveling wave solutions (like one-soliton solutions) of \rf{eqofmotgenv}. In fact, such conclusion applies for any functional of the scalar field $\vp$ and its derivatives, which is a tensor under the Lorentz group. 

The one-solitons we treat in this paper  are localized solutions in the sense that the field $\vp$ have non-vanishing space-time derivatives only in a  small region of space. Therefore, the integrand in the definition \rf{quasiconserved} of the anomaly $\beta^{(N)}$, is non-vanishing only in such a small region of space, i.e., it gets exponentially suppressed outside such region. In addition, the one-solitons interact with each other by short range interactions. So, a two-soliton solution for the case when the two one-solitons are far apart should be just the superposition of the one-soliton solutions. Therefore, the anomaly $\beta^{(N)}$ evaluated on a two-soliton solution when the one-solitons are far apart should vanish, because it reduces to the sum of the anomalies of the two one-solitons, which by the argument above vanish.
Consequently one should expect the anomaly $\beta^{(N)}$ to be non-vanishing only when the solitons are close together and interacting. That is in fact what we observe in our numerical simulations described in section \ref{sec:simulations} . We do not have yet a good understanding of non-linear dynamics governing the behavior of the charges and anomalies when the solitons interact with each other. That is a crucial issue to be understood and is at the heart of our working definition of the concept of quasi-integrability. What is clear so far is that  special properties of the solutions under a space-time parity transformation play an important role in all that. It is not clear however if such properties are the causes or consequences of the quasi-integrability. Let us explain how it works. 

 Let us then consider a (two-soliton like) solution of the equations of motion  
\rf{eqofmotgenv}, and a space-time  parity transformation
\be
P:\qquad\qquad \({\tilde x}\, , \, {\tilde t}\)\rightarrow \(-{\tilde x}\, , \,- {\tilde t}\) \qquad\qquad 
{\tilde x}=x-x_{\Delta} \qquad {\tilde t}=t-t_{\Delta}
\lab{paritydef}
\ee
where the values of $x_{\Delta}$ and $t_{\Delta}$ depend upon the parameters of that particular solution. There are two important classes of two-soliton solutions according to the way they behave under the parity transformation. The first one is that where the two-soliton solution is invariant under the parity, i.e. $P\(\vp\)=\vp$. From the arguments below \rf{alphatransform} we concluded that each term in $\alpha^{(N)}$ must contain $(N-1)$ $x_{-}$-derivatives of $\vp$. From \rf{Ndefinaplus} we have that the integer $N$ is of the form $N=6\,l\pm1$, with $l$ any integer. Therefore,  each term in $\alpha^{(N)}$ contains an even number of $x_{-}$-derivatives of $\vp$, and so it is invariant under the parity. Therefore, $P\(\partial_{-}\varphi\, X\,\alpha^{(N)}\)=-\partial_{-}\varphi\, X\,\alpha^{(N)}$. Consequently 
\be
\int_{-{\tilde t}_0}^{{\tilde t}_0} d{\tilde t}\,\int_{-{\tilde x}_0}^{{\tilde x}_0} d{\tilde x}\,
\partial_{-}\varphi\, X\,\alpha^{(N)}=0
\lab{notnicecancel}
\ee
where ${\tilde x}_0$ and ${\tilde t}_0$ are any chosen values of ${\tilde x}$ and ${\tilde t}$ respectively, i.e. we are integrating on a rectangle with center in $\(x_{\Delta}\, , \, t_{\Delta}\)$ 
(see \rf{paritydef}). Therefore, the  charges evaluates on such two-soliton solutions satisfy the mirror like symmetry
\be
Q^{(N)}\({\tilde t}_0\)=Q^{(N)}\(-{\tilde t}_0\)
\lab{mirrorsymmetrynotnice}
\ee
for any ${\tilde t}_0$. The asymptotic conservation of the charges is therefore a particular case of such mirror symmetry, corresponding to the case where ${\tilde t}_0\ra \infty$. 

The physically important two-soliton solutions however, are not invariant under the parity transformation. The  complex scalar field $\vp$, evaluated on such two-soliton solutions, satisfy the property 
\be
P\(\varphi\)=\varphi^*
\lab{phiunderp}
\ee

We are assuming that the potentials $V$ of our theories satisfy the property \rf{conjugatepot}, i.e., the potential depends only on $\vp$ and not on its complex conjugated $\vp^*$, and the complex conjugated of $V$ is obtained just by the replacement $\vp\ra \vp^*$. Therefore the anomaly $X$, defined in \rf{xdef}, also satisfy the same property, i.e.,
\be
X^*=X\(\vp\ra \vpc\)
\lab{conjugatex}
\ee
In addition, the Lax potentials $A_{\pm}$, defined in \rf{deformedpot}, do not involve complex parameters, and according to the appendix \ref{app:kacmoody} the structure constants of the loop algebra $A_2^{(2)}$, in the basis $b_N$ and $F_n$, are all real. Consequently, the complex conjugate of the charges $Q^{(N)}$ and of the quantities $\alpha^{(N)}$, defined in \rf{chargesdef} and \rf{rotatef1} respectively, are also obtained just by the replacement $\vp\ra\vp^*$, i.e.
\be
\(Q^{(N)}\)^*=Q^{(N)}\(\vp\ra\vp^*\)\; ;\qquad\qquad\qquad 
\(\alpha^{(N)}\)^*=\alpha^{(N)}\(\vp\ra\vp^*\)
\ee
As we have argued above, each term in the quantity  $\alpha^{(N)}$ contains an even number of $x_{-}$-derivatives of $\vp$, and so  for the two-soliton solutions satisfying \rf{phiunderp}, one gets that 
\be
P\(\partial_{-}\vp\,X\,\alpha^{(N)}\)= -\partial_{-}\vp^*\,X^*\, \(\alpha^{(N)}\)^*
\lab{anomalyunderparity}
\ee
Consequently 
\be
\int_{-{\tilde t}_0}^{{\tilde t}_0} d{\tilde t}\,\int_{-{\tilde x}_0}^{{\tilde x}_0} d{\tilde x}\,
\(\partial_{-}\varphi\, X\,\alpha^{(N)}+\partial_{-}\varphi^*\, X^*\,\(\alpha^{(N)}\)^*\)
=0
\ee
where, like in \rf{notnicecancel}, we are integrating on a rectangle with center in $\(x_{\Delta}\, , \, t_{\Delta}\)$ (see \rf{paritydef}). Therefore, for the two-soliton solutions satisfying \rf{phiunderp}, the real part of the charges $Q^{(N)}$ satisfy the mirror like symmetry
\be
\(Q^{(N)}+\(Q^{(N)}\)^*\)\({\tilde t}_0\)=\(Q^{(N)}+\(Q^{(N)}\)^*\)\(-{\tilde t}_0\)
\lab{mirrorsymmetry}
\ee
for any ${\tilde t}_0$. Consequently, in the limit ${\tilde t}_0\ra \infty$, we get that the real parts of the charges are asymptotically conserved, i.e., they have the same values before and after the scattering of the solitons.

\section{The interplay between parity and the perturbative expansion}
\label{sec:parityversusanomaly}
\setcounter{equation}{0}

Let us consider potentials $V$ that are perturbations of the Bullough-Dodd  potential, in the sense that they depend upon a parameter $\ve$ such that they become the Bullough-Dodd  potential for $\ve=0$. The potential \rf{deformpot} is an example of it. We shall expand the solutions in a power series in the parameter $\ve$ as
\be
\vp=\vp_0+\ve\,\vp_1+\ve^2\,\vp_2+\ldots
\lab{perturbexpand}
\ee
Therefore the potential depends explicitly upon $\ve$ and also implicitly through $\vp$, and so we have the expansion
\br
\frac{\partial \, V}{\partial \, \vp}&=& \frac{\partial \, V}{\partial \, \vp}\mid_{\ve=0}+ \,\ve\,\left[\frac{\partial^2 \, V}{\partial \,\ve\,\partial \, \vp}
+ \frac{\partial^2 \, V}{\partial \, \vp^2} \,\frac{\partial \, \vp}{\partial \,\ve} \right]_{\ve=0}
\lab{expandforce}\\
&+&\frac{\ve^2}{2}\,\left[\frac{\partial^3 \, V}{\partial \,\ve^2\,\partial \, \vp}
+ 2\,\frac{\partial^3 \, V}{\partial\,\ve\partial \, \vp^2} \,\frac{\partial \, \vp}{\partial \,\ve} + \frac{\partial^2 \, V}{\partial \, \vp^2} \,\frac{\partial^2 \, \vp}{\partial \,\ve^2} 
+ \frac{\partial^3 \, V}{\partial \, \vp^3} \,\(\frac{\partial \, \vp}{\partial \,\ve} \)^2\right]_{\ve=0}+O\(\ve^3\)
\nonumber
\er 
From the equation of motion \rf{eqofmotgenv} we then get the equations for the components of the field $\vp$ in the expansion \rf{perturbexpand}, as ($\partial^2\equiv \partial^2_t-\partial^2_x$)
\br
\partial^2\vp_0 +   e^{\vp_0}-e^{-2\,\vp_0} &=&0
\lab{zeroordereq}\\
\partial^2\vp_n +   \frac{\partial^2 \, V}{\partial \, \vp^2}\mid_{\ve=0}\, \vp_n &=&
f_n \qquad\qquad \qquad i=1, 2, 3\ldots
\lab{nordereq}
\er
with
\br
f_1&=& -\frac{\partial^2 \, V}{\partial \,\ve\,\partial \, \vp}\mid_{\ve=0}
\lab{fndef}\\
f_2&=& -\frac{1}{2}\left[\frac{\partial^3 \, V}{\partial \,\ve^2\,\partial \, \vp}\mid_{\ve=0}
+ 2\,\frac{\partial^3 \, V}{\partial\,\ve\partial \, \vp^2}\mid_{\ve=0} \,\vp_1 
+ \frac{\partial^3 \, V}{\partial \, \vp^3}\mid_{\ve=0} \,\vp_1^2\right]
\nonumber
\er
and so on. 

Let us split the fields into the even and odd parts under the parity transformation \rf{paritydef}, and into their real and imaginary parts as well
\be
\vp_n^{\pm}\equiv \frac{1}{2}\(1\pm P\)\; ;\qquad\qquad\qquad
\vp_n^{R}\equiv \frac{1}{2}\(1+ C\)\; ; \qquad\quad
\vp_n^{I}\equiv \frac{1}{2\,i}\(1- C\)
\ee
with $C$ being the complex conjugation operation. By splitting the zero order equation \rf{zeroordereq} into its even and odd parts under the parity one gets
\be
\partial^2\vp_0^{\pm} + \frac{1}{2}\,\left[  e^{\vp_0^{+}+\vp_0^{-}}
\pm  e^{\vp_0^{+}-\vp_0^{-}}
-\(e^{-2\,\(\vp_0^{+}+\vp_0^{-}\)} \pm e^{-2\,\(\vp_0^{+}-\vp_0^{-}\)}\)\right]=0
\ee
Note therefore that one should not expect non-trivial solutions for the case $\vp_0^{+}=0$, since $\vp_0^{-}$ would have to assume some very special (vacuum) constant values. On the other hand the case $\vp_0^{-}=0$ can lead to non-trivial solutions. Following \rf{phiunderp} we shall therefore consider solutions of the pure Bullogh-Dodd equation \rf{zeroordereq} which satisfies
\be
P\(\vp_0\)=\vp_0^* \; ;\qquad\qquad {\rm i.e.}\qquad \qquad 
P\(\vp_0^R\)=\vp_0^R\qquad P\(\vp_0^I\)=-\vp_0^I
\lab{phi0underp}
\ee
Note that $f_1$ given in \rf{fndef} is a function of the order zero field $\vp_0$ only. But since we are assuming that the potential satisfies the property \rf{conjugatepot}, it follows from \rf{phi0underp} that the action of the parity $P$ on $f_1$ is the same as the action of the complex conjugation $C$. Therefore
\be
\frac{1}{2}\,\(1\pm P\)\,\frac{1}{2}\,\(1\pm C\)\,f_1=f_1^{\pm}\; ;
\qquad\qquad\qquad
\frac{1}{2}\,\(1\pm P\)\,\frac{1}{2}\,\(1\mp C\)\,f_1=0
\ee
But since $\frac{\partial^2 \, V}{\partial \, \vp^2}\mid_{\ve=0}$ is also a function of $\vp_0$ only, it follows from the same reasoning that
\br
\frac{1}{2}\,\(1\pm P\)\,\frac{1}{2}\,\(1\pm C\)\,\frac{\partial^2 \, V}{\partial \, \vp^2}\mid_{\ve=0}=\(\frac{\partial^2 \, V}{\partial \, \vp^2}\mid_{\ve=0}\)^{\pm}\; ;
\qquad\quad
\frac{1}{2}\,\(1\pm P\)\,\frac{1}{2}\,\(1\mp C\)\,\frac{\partial^2 \, V}{\partial \, \vp^2}\mid_{\ve=0}=0
\nonumber
\er
Therefore splitting the equation \rf{nordereq} for $f_1$ into its even and odd parts under $P$ and its real and imaginary components one gets
\br
\partial^2\vp_1^{R,+} 
+   \(\frac{\partial^2 \, V}{\partial \, \vp^2}\mid_{\ve=0}\)^{+}\, \vp_1^{R,+}
+i\,  \(\frac{\partial^2 \, V}{\partial \, \vp^2}\mid_{\ve=0}\)^{-}\, \vp_1^{I,-}&=& f_1^{+}
\nonumber\\
\partial^2\vp_1^{I,-} 
+   \(\frac{\partial^2 \, V}{\partial \, \vp^2}\mid_{\ve=0}\)^{+}\, \vp_1^{I,-}
-i\,  \(\frac{\partial^2 \, V}{\partial \, \vp^2}\mid_{\ve=0}\)^{-}\, \vp_1^{R,+}&=& -i\,f_1^{-}
\nonumber\\
\partial^2\vp_1^{I,+} 
+   \(\frac{\partial^2 \, V}{\partial \, \vp^2}\mid_{\ve=0}\)^{+}\, \vp_1^{I,+}
-i\, \(\frac{\partial^2 \, V}{\partial \, \vp^2}\mid_{\ve=0}\)^{-}\, \vp_1^{R,-}&=&0
\lab{phi1eqs}\\
\partial^2\vp_1^{R,-} 
+   \(\frac{\partial^2 \, V}{\partial \, \vp^2}\mid_{\ve=0}\)^{+}\, \vp_1^{R,-}
+i\, \(\frac{\partial^2 \, V}{\partial \, \vp^2}\mid_{\ve=0}\)^{-}\, \vp_1^{I,+}&=&0
\nonumber
\er
So, the pair of fields $\(\vp_1^{R,+}\, , \, \vp_1^{I,-}\)$ satisfy a pair of linear non-homogeneous equations, and the pair of fields $\(\vp_1^{R,-}\, , \, \vp_1^{I,+}\)$ satisfy a pair of linear homogeneous equations. In addition, the two pairs of equations are decoupled. Therefore, $\(\vp_1^{R,-}\, , \, \vp_1^{I,+}\)=0$ is a solutions of such equations, but the pair $\(\vp_1^{R,+}\, , \, \vp_1^{I,-}\)$ can never vanish. If one has a given solution $\(\vp_1^{R}\, , \, \vp_1^{I}\)$ of the equations above, then the configuration $\({\tilde \vp}_1^{R,+}\, , \, {\tilde \vp}_1^{I,-}\)\equiv \(\vp_1^{R}\, , \, \vp_1^{I}\)- \(\vp_1^{R,-}\, , \, \vp_1^{I,+}\)$, is also a solution. Therefore, given a solution one can always make its real part even under $P$, and its imaginary part odd under $P$, i.e. one can always choose the first order solution to satisfy
\be
P\(\vp_1\)= \vp_1^*
\ee 
The quantity $f_2$ given in \rf{fndef} is a function of $\vp_0$ and $\vp_1$ only. If the potential $V$ satisfy the property \rf{conjugatepot} then it follows, by the same arguments used above, that 
\be
\frac{1}{2}\,\(1\pm P\)\,\frac{1}{2}\,\(1\pm C\)\,f_2=f_2^{\pm}\; ;
\qquad\qquad\qquad
\frac{1}{2}\,\(1\pm P\)\,\frac{1}{2}\,\(1\mp C\)\,f_2=0
\ee
Consequently the pairs of fields $\(\vp_2^{R,+}\, , \, \vp_2^{I,-}\)$ and $\(\vp_2^{R,-}\, , \, \vp_2^{I,+}\)$, satisfy equations identical to \rf{phi1eqs} with $f_1$ replaced by $f_2$. Therefore, by same arguments as above, one can always choose the second order solution to satisfy 
\be
P\(\vp_2\)= \vp_2^*
\ee 
Continuing with such process order by order, one concludes that is always possible to choose a solution, as long as the perturbative series converge, that satisfies the property \rf{phiunderp} which was used to prove the mirror symmetry property \rf{mirrorsymmetry} satisfied by the real part of the charges 
$Q^{(N)}$. 

The conclusion is that the deformations of the Bulough-Dodd model we are considering seems to possess a subclass of solutions that present an infinite number of real charges which are quasi-conserved. By quasi-conserved we mean charges satisfying the mirror symmetry \rf{mirrorsymmetry}. In the case of two-soliton like solutions such properties imply that the infinity of real charges preserve the same values, after the scattering of the solitons, that they had prior the scattering, even though during the scattering process itself they may vary considerably. It is that sub-sector of the model, consisting of solutions satisfying 
\rf{phiunderp}, that we call a {\em quasi-integrable} theory. 

\section{The Hirota's solutions of the Bullough-Dood model}
\label{sec:hirota}
\setcounter{equation}{0}

We now show that the pure Bullough-Dood model possesses soliton solutions satisfying the property 
\rf{phi0underp}. We shall use the Hirota's method to construct such solutions, and so we introduce the $\tau$-functions as 
\be
\vp_0=\ln\frac{\tau_0}{\tau_1}
\lab{taudef}
\ee
One can check that if such $\tau$-functions satisfy the Hirota's equations
\br
\tau_0 \partial_{+}\partial_{-}\tau_0 - \partial_{+} \tau_0
\,\partial_{-}\tau_0 &=& \tau_1^2-\tau_0^2\nonu\\ 
\tau_1 \partial_{+}\partial_{-}\tau_1 - \partial_{+} \tau_1
\,\partial_{-}\tau_1 &=& \tau_0 \,\tau_1-\tau_1^2 
\lab{hirotaeqcbd}
\er
then the zero order field $\vp_0$, given in \rf{taudef}, satisfy the  Bullough-Dood equation  
\rf{zeroordereq}.

\subsection{One-soliton solutions}

The one-soliton  solutions of the pure Bullough-Dood model correspond to  the following solutions of the Hirota's equations  \rf{hirotaeqcbd} 
\br
\tau_0= 1 -4\, a\, e^{\Gamma} + a^2\,e^{2\,\Gamma}\; ; \qquad\qquad\qquad 
\tau_1= \(1 + a\, e^{\Gamma}\)^2 
\lab{onesolbd}
\er
with 
\be
\Gamma = \sqrt{3}\,\( z\, x_{+}-\frac{x_{-}}{z}\)
\ee
where $z$ and $a$ are complex parameters. The one-soliton solutions leading to the quasi-integrable sector of the deformed theories we are interested in, correspond to the cases where $z$ is real. Then we parameterize it as
$z=e^{-\alpha}$, and define  $ v= \tanh\alpha$, where $v$ is the soliton velocity and so $\alpha$ is the rapidity.  We now write $a=e^{i\, \xi}\, e^{\frac{-\sqrt{3}\, x^{(0)}}{\sqrt{1-v^2}}}$ and define 
$a\, e^{\Gamma}\equiv e^{W}$, with 
\be 
W\equiv \frac{\sqrt{3}}{\sqrt{1-v^2}}\,\(x-v\,t-x^{(0)}\)+i\, \xi
\lab{wonesoliton}
\ee
Therefore, we get from \rf{onesolbd}
\be
\frac{\tau_0}{\tau_1}= \frac{\cosh W -2}{\cosh W +1}
\lab{niceonesol}
\ee
Note that if the phase $\xi$ vanishes, then $\frac{\tau_0}{\tau_1}$ vanishes whenever $\cosh W=2$, and that corresponds to a singularity in the solution for $\vp_0$. Therefore, we shall restrict to the cases where $\xi\neq 0$.  We have that
\be
\cosh W= \cos \xi\, \cosh \(\frac{\sqrt{3}}{\sqrt{1-v^2}}\,\(x-v\,t-x^{(0)}\)\)   
+i\, \sin \xi\,\sinh\(\frac{\sqrt{3}}{\sqrt{1-v^2}}\,\(x-v\,t-x^{(0)}\)\)  
\ee
Therefore, under the space-time parity transformation
\be
\(x-x^{(0)}\)\ra -\(x-x^{(0)}\) \qquad\qquad \qquad\qquad t\ra -t
\ee 
one gets that
\be
\cosh W \ra \cosh W^* \qquad\qquad\qquad \mbox{\rm and so}\qquad\qquad  \vp_0\ra \vp_0^*
\ee
Therefore, the one-soliton solutions \rf{niceonesol} satisfy the property \rf{phi0underp}. 
By the arguments of section \ref{sec:parityversusanomaly} such one-soliton solution of the pure Bullough-Dodd model can serve as a seed to construct, by a perturbative approach,  one-soliton solutions of the deformed theory that satisfy the property  \rf{phiunderp}. 

However, by the arguments presented below \rf{tensoranomaly}, for any traveling wave solution, and so for any one-soliton solution, the charges $Q^{(N)}$, given in \rf{chargesdef}, are not only quasi-conserved but exactly conserved. 

\subsection{Two soliton solution}

The solutions on the two-soliton sector of the pure Bullough-Dood model are obtained by  solving \rf{hirotaeqcbd} by the Hirota's method, and the result is 
\br
\tau_0&=& 1 -4\, a_1\, e^{\Gamma_1}-4\, a_2\, e^{\Gamma_2}  
+a_1^2\,e^{2\,\Gamma_1}
+a_2^2\,e^{2\,\Gamma_2}\nonu\\
&+&8\,a_1\,a_2\,\frac{2\,z_1^4-z_1^2\,z_2^2+2\,z_2^4}{
\(z_1+z_2\)^2\(z_1^2+z_1\,z_2+z_2^2\)} \, e^{\Gamma_1+\Gamma_2} 
\nonu\\
&-&4\,a_1^2\,a_2\,
\frac{\(z_1-z_2\)^2\(z_1^2-z_1\,z_2+z_2^2\)}{
\(z_1+z_2\)^2\(z_1^2+z_1\,z_2+z_2^2\)} \, e^{2\,\Gamma_1+\Gamma_2}
\nonu\\
&-&4\,a_1\,a_2^2\,
\frac{\(z_1-z_2\)^2\(z_1^2-z_1\,z_2+z_2^2\)}{
\(z_1+z_2\)^2\(z_1^2+z_1\,z_2+z_2^2\)} \, e^{\Gamma_1+2\,\Gamma_2}
\nonu\\
\nonu\\
&+&\,a_1^2\,a_2^2\,
\frac{\(z_1-z_2\)^4\(z_1^2-z_1\,z_2+z_2^2\)^2}{
\(z_1+z_2\)^4\(z_1^2+z_1\,z_2+z_2^2\)^2} \, e^{2\,\Gamma_1+2\Gamma_2}
\nonu\\
\tau_1&=& 1 + 2\, a_1\, e^{\Gamma_1} + 2\, a_2\, e^{\Gamma_2}
+a_1^2\,e^{2\,\Gamma_1}
+a_2^2\,e^{2\,\Gamma_2}\nonu\\
&+&4\, a_1\,a_2\,\frac{z_1^4+4\,z_1^2\,z_2^2+z_2^4}{
\(z_1+z_2\)^2\(z_1^2+z_1\,z_2+z_2^2\)}  \, e^{\Gamma_1+\Gamma_2}
\nonu\\
&+&2\,a_1^2\,a_2\,
\frac{\(z_1-z_2\)^2\(z_1^2-z_1\,z_2+z_2^2\)}{
\(z_1+z_2\)^2\(z_1^2+z_1\,z_2+z_2^2\)} \, e^{2\,\Gamma_1+\Gamma_2}
\nonu\\
&+&2\,a_1\,a_2^2\,
\frac{\(z_1-z_2\)^2\(z_1^2-z_1\,z_2+z_2^2\)}{
\(z_1+z_2\)^2\(z_1^2+z_1\,z_2+z_2^2\)} \, e^{\Gamma_1+2\,\Gamma_2}
\nonu\\
&+&\,a_1^2\,a_2^2\,
\frac{\(z_1-z_2\)^4\(z_1^2-z_1\,z_2+z_2^2\)^2}{
\(z_1+z_2\)^4\(z_1^2+z_1\,z_2+z_2^2\)^2} \, e^{2\,\Gamma_1+2\Gamma_2}
\lab{twosolcbd}
\er
where 
\be
\Gamma_i = \sqrt{3}\,\( z_i\, x_{+}-\frac{x_{-}}{z_i}\) \qquad\qquad \qquad \qquad i=1,2
\lab{gammadef}
\ee
and where $z_i$ and $a_i$, $i=1,2$, are complex parameters. As in the case of the one-soliton solutions, we shall deal with the two-soliton solutions where $z_i$ are real,   and parameterize the solutions with six real parameters $v_i$, $\xi_i$ and $ x^{(0)}_i$, $i=1,2$,  as 
\be
z_i=e^{-\alpha_i} \; ; \qquad\qquad \quad
a_i=e^{i\, \xi_i}\, e^{\frac{-\sqrt{3}\, x^{(0)}_i}{\sqrt{1-v_i^2}}} \; ; \qquad\qquad\quad
v_i=\tanh \alpha_i
\ee
Similarly to what we have done in the case of one-soliton let us define $a_i\,e^{\Gamma_i}\equiv e^{W_i}$, with
\be
W_i\equiv \sqrt{3}\,\frac{\(x-v_i\,t-x^{(0)}_i\)}{\sqrt{1-v_i^2}}+i\,\xi_i\qquad\qquad \qquad \qquad i=1,2
\lab{widef}
\ee
In addition we introduce the real quantities $\Delta$, $c_0$ and $c_1$ as
\br
e^{\Delta\(v_i\)}&\equiv &
\frac{\(z_1-z_2\)^2\(z_1^2-z_1\,z_2+z_2^2\)}{
\(z_1+z_2\)^2\(z_1^2+z_1\,z_2+z_2^2\)}
\nonumber\\
c_0\(v_i\)&\equiv&
\frac{2\,z_1^4-z_1^2\,z_2^2+2\,z_2^4}{
\(z_1-z_2\)^2\(z_1^2-z_1\,z_2+z_2^2\)}
\lab{deltadef}\\
c_1\(v_i\)&\equiv&
\frac{z_1^4+4\,z_1^2\,z_2^2+z_2^4}{
\(z_1-z_2\)^2\(z_1^2-z_1\,z_2+z_2^2\)}
\nonumber
\er
Therefore, the Hirota's $\tau$-functions become
\br
\tau_0&=& 1- 4\, e^{W_1}- 4\, e^{W_2} + e^{2\,W_1}+ e^{2\,W_2} 
+ e^{W_1+W_2+\Delta}\,\left[8\, c_0- 4\, e^{W_1}- 4\, e^{W_2} + e^{W_1+W_2+\Delta}\right]
\nonumber\\
\tau_1&=& 1+2\, e^{W_1}+2\, e^{W_2} + e^{2\,W_1}+ e^{2\,W_2} 
+ e^{W_1+W_2+\Delta}\,\left[4\, c_1+2\, e^{W_1}+2\, e^{W_2} + e^{W_1+W_2+\Delta}\right]
\nonumber
\er
We then introduce the new space-time coordinates as 
\br
\chi_{+}&\equiv& \sqrt{3}\left[\frac{\(x-v_1\,t-x^{(0)}_1\)}{\sqrt{1-v_1^2}}
+\frac{\(x-v_2\,t-x^{(0)}_2\)}{\sqrt{1-v_2^2}}\right]+\Delta 
\nonumber\\ 
\chi_{-}&\equiv& \sqrt{3}\left[\frac{\(x-v_1\,t-x^{(0)}_1\)}{\sqrt{1-v_1^2}}
-\frac{\(x-v_2\,t-x^{(0)}_2\)}{\sqrt{1-v_2^2}}\right]
\lab{xpmdef}
\er
Note that as long as $v_1\neq v_2$, $\chi_{\pm}$ are independent coordinates. In fact, one can check if $v_1=v_2$ the two-soliton solution \rf{twosolcbd} reduces to a one-soliton solution \rf{onesolbd} (taking $z_1=z_2\equiv z$ and $a_1+a_2\equiv a$) . In addition we denote
\be
W_{+}\equiv W_1+W_2+\Delta =\chi_{+}+i\, \(\xi_1+\xi_2\)\; ;
\qquad\quad
W_{-}\equiv W_1-W_2= \chi_{-}+i\, \(\xi_1-\xi_2\)
\lab{wtwosoliton}
\ee
Therefore, by factoring out from $\tau_0$ and $\tau_1$, the term $2\,e^{W_1+W_2+\Delta}$, one gets that 
\br
\frac{\tau_0}{\tau_1}=\frac{\cosh W_{+}+4\, c_0- 8\,e^{-\frac{1}{2}\Delta}\,\cosh\(\frac{W_{+}}{2}\)\,\cosh\(\frac{W_{-}}{2}\) + e^{-\Delta}\, \cosh W_{-}}{\cosh W_{+}+2\, c_1+4\,e^{-\frac{1}{2}\Delta}\,\cosh\(\frac{W_{+}}{2}\)\,\cosh\(\frac{W_{-}}{2}\)+ e^{-\Delta}\, \cosh W_{-}}
\lab{finalratiotau0tau1}
\er
Note that under  the space-time parity transformation
\be
P:\qquad\qquad\qquad \(\chi_{+}\, ,\, \chi_{-}\)\rightarrow \(-\chi_{+}\, ,\, -\chi_{-}\)
\lab{paritydef2sol}
\ee
one gets that
\be
P\(W_{\pm}\)=-W_{\pm}^*
\ee
and consequently
\be
P\(\frac{\tau_0}{\tau_1}\)=\frac{\tau_0^*}{\tau_1^*}
\ee
Therefore, from \rf{taudef} one observes that such two-soliton solutions of the pure Bullough-Dood model satisfy the property  \rf{phi0underp}. By the arguments of section \ref{sec:parityversusanomaly} such two-soliton solution  can serve as a seed to construct, by a perturbative approach,  two-soliton solutions of the deformed theory that satisfy the property  \rf{phiunderp}.

\section{The numerical simulations}
\label{sec:simulations}
\setcounter{equation}{0}

We performed various simulations of the proposed deformed Bullough-Dodd model
to check the ideas behind the concept of quasi-integrability, specially the behavior of the anomalies 
\rf{quasiconserved} and \rf{intanomaly} during the scattering of two solitons.  To do so we had first to generate static solutions
to the deformed equations of motion. Then we stitched two of them together to perform
a kink-kink colision and analysed the results.\\

In all our numerical work the time evolution was simulated by the fourth order
Runge-Kutta method. After several lattice sizes were tested we found that a
lattice extending from $-25$ to $+25$ was enough to obtain reliable results.
We also found that a grid spacing of $0.01$ in space and of $0.003$ in time
discretization provided enough accuracy. Our initial condition
for the kink-kink scattering is not exact, thus radiation
was generated. This radiation was absorbed at the boundaries
through a viscous force acting from $x=-25$ to $x=-22$ and from $x=+22$ to $x=+25$.
Such a boundary condition effectively simulates an infinite grid.

\subsection{Generating the initial condition}
To obtain static solutions of \rf{deformeqom}
we used a minimization method that consists in solving
\begin{equation}
\lab{static_eq}
\partial_\tau\varphi-\partial_x^2\varphi+e^\varphi-e^{-(2+\epsilon)\varphi}=0.
\end{equation}
This equation is an evolution in the fake time $\tau$.
The static solutions of \rf{static_eq} and those
of \rf{deformeqom} are the same, but only the former
is dissipative. The dissipative nature and the absence of sources
means that any initial condition will lead to a static solution in the infinite-time limit.
We used the exact solutions of the Bullough-Dodd model as seeds to this method, and labeled
the results with the original parameters -- i.e. seeding the kink obtained with $\xi=0.9$ from
the original model ($\epsilon=0$) into \rf{static_eq} results in the deformed kink with $\xi=0.9$.

\begin{figure}
\end{figure}

Once we stitch two kinks together we can use two facts to
generate the complete initial condition: first, the two kinks (let's call them the left kink and the right kink)
are far from each other and can be treated as free kinks;
second, each independent kinks have Lorentz invariace so
\begin{equation}
\lab{initial_conditions}
\varphi_i=\varphi_i (x-vt) \Rightarrow \frac{\partial \varphi_i}{\partial x} = -\frac{1}{v}\frac{\partial \varphi_i}{\partial t}
\end{equation}
where $\varphi_i$ denotes any of the two kinks in the absence of the other.
We use $\varphi_i$ as the left kink in \rf{initial_conditions} from the grid's left edge
to position $x=0$, and as the right kink for the rest of the grid. This also introduces
some perturbation with respect to the exact two kink solution, but one
that is small for small colision velocities.

\subsection{Kink-kink interaction}
Once we have generated our initial conditions we simulated the colision between two kinks,
and calculated the first non-trivial anomaly  $\beta^{(N)}$ and  integrated anomaly $\gamma^{(N)}$, for $N=5$, given in \rf{quasiconserved} and \rf{intanomaly} respectively. Using \rf{xdef}, 
\rf{anomalyfinal}, \rf{quasiconserved} and \rf{intanomaly}, and for the potential \rf{deformpot} one gets that  (with the choice $m=0$ in \rf{xdef})
\be
\beta^{(5)}=\frac{\ve\,\(\ve+3\)}{2\(2+\ve\)}\int_{-\infty}^{\infty}dx\, \partial_{-}\vp e^{-\(2+\ve\)\vp}\left[4  \(\partial_{-}\varphi\)^2 \partial_{-}^2\varphi-4 
   \(\partial_{-}^2\varphi\)^2-2 \, \partial_{-}\varphi
   \partial_{-}^3\varphi-2 \, \partial_{-}^4\varphi\right]
   \lab{beta5}
   \ee
   and
\be
\gamma^{(5)}=\frac{\ve\,\(\ve+3\)}{2\(2+\ve\)}\int_{-\infty}^{t}\int_{-\infty}^{\infty}dx\, \partial_{-}\vp e^{-\(2+\ve\)\vp}\left[4  \(\partial_{-}\varphi\)^2 \partial_{-}^2\varphi-4 
   \(\partial_{-}^2\varphi\)^2-2 \, \partial_{-}\varphi
   \partial_{-}^3\varphi-2 \, \partial_{-}^4\varphi\right]  
   \lab{gamma5}
   \ee 
Note that  $\ve=-3$ is special because the potential \rf{deformpot} vanishes in that case. 

We have performed a large number of  simulations of the scattering of two kinks for various values of the deformation parameter $\ve$ and for the phases $\xi_i$, $i=1,2$, which appear in the exact soliton solutions (see \rf{wonesoliton} and \rf{wtwosoliton}). Note that such phases are parameters of the exact soliton solutions but we do not know how the deformed solitons depend upon them for $\ve\neq 0$. However, as explained above, we have constructed the static kink numerically (with $\ve\neq 0$) from an exact kink (with $\ve=0$) for a given choice of the phase $\xi_i$. So, we still use the phases $\xi_i$ to label the deformed kink. 

The results for the scattering of two kinks for various choices of the parameters $\ve$ and  $\xi_i$, $i=1,2$, were qualitatively the same, and we show in the figures \ref{fig:10909}, \ref{fig:10914}, \ref{fig:10920}, \ref{fig:20920} and \ref{fig:m10920}, some representative cases.  On those figures we show on the top plots,  the real and imaginary parts of the anomaly $\beta^{(5)}$, given in \rf{beta5}, and on the bottom plots the  real and imaginary parts of the integrated anomaly $\gamma^{(5)}$, given in \rf{gamma5}. 

Note that, in the cases shown in figures \ref{fig:10909}-\ref{fig:m10920},  the real part of the integrated anomaly  $\gamma^{(5)}$ is symmetric under reflection around a point close to $t=0$. From \rf{intanomaly} one has 
\be
\gamma^{(N)}\({\tilde t}_0\)- \gamma^{(N)}\(-{\tilde t}_0\)= Q^{(N)}\({\tilde t}_0\)-Q^{(N)}\(-{\tilde t}_0\)= \frac{1}{2}\, \int_{-{\tilde t}_0}^{{\tilde t}_0} dt \,\int_{-\infty}^{\infty} dx\,\partial_{-}\varphi\, X\,\alpha^{(N)}
\lab{intanomaly2}
\ee
According to the numerical simulations, the real part of $\left[\gamma^{(5)}\({\tilde t}_0\)- \gamma^{(5)}\(-{\tilde t}_0\)\right]$ vanishes, and so one gets a numerical confirmation of \rf{mirrorsymmetry}, for $N=5$ at least. Remember from \rf{paritydef} that ${\tilde t}=t-t_{\Delta}$, and for the simulations shown in figures \ref{fig:10909}-\ref{fig:m10920}, one has $t_{\Delta}$ close to zero.

The case $\ve=-1$, shown in \ref{fig:m10920}, has an asymmetry however, in the oscillations of the anomaly, which is perhaps due to numerical errors. Therefore, our numerical results confirm the existence of a mirror symmetry given in   \rf{mirrorsymmetry}, and obtained from the analytical calculations based on the parity properties of the solutions. In addition, such results imply that real part of the charge is asymptotically conserved, i.e. ${\rm Re}\, Q^{(5)}\(t=-\infty\)= {\rm Re}\, Q^{(5)}\(t=\infty\)$. The plots of the imaginary part of the integrated anomaly  $\gamma^{(5)}$ show that the imaginary part of the charge $Q^{(5)}$  is not asymptotically conserved, and so it is in agreement with our analytical calculations that lead to the asymptotic conservation of the real part of the charge only. 

 \section{Conclusions}
 \label{sec:conclusions}
\setcounter{equation}{0}
 
 We have considered deformations of the exactly integrable Bullough-Dodd model defined by the Lagrangians \rf{deformbdlag} and potentials \rf{deformpot}, involving a deformation parameter $\ve$. We have shown that such deformed theories possess sectors where the soliton-like solutions present properties very similar to solitons in exactly integrable field theories. They possess an infinite set of quantities which are exactly conserved in time for one-soliton type solutions, i.e. solutions traveling with a constant speed, and are asymptotically conserved for two-soliton type solutions.  By asymptotically conserved we mean quantities which do vary in time during the scattering process of two one-solitons but that return, in the distant future  (after the collision), to the values they had in the distant past (before the collision). Since what matters in a scattering process are the asymptotic states, such behavior is effectively what we observe in scattering of solitons in integrable field theories, and that is why we call such theories {\em quasi-integrable}. The mechanism behind the asymptotic conservation of such an infinity of charges is not well understood yet, but in all examples where it has been observed so far 
 \cite{quasi1,quasi2,quasi3,quasi4}, the two-soliton type solutions present special properties under a space-time parity transformation. In the case of the deformations considered in this paper, the complex scalar field $\vp$ of the theory, when evaluated on the two-soliton solutions, is mapped into its complex conjugate under the parity transformation (see \rf{phiunderp}).  It is worth mentioning that the point around which space and time are reversed, under the parity transformation, depend upon the parameters of the solution under consideration. 
 
 The infinity of asymptotically conserved charges was constructed using techniques of integrable field theories based on an anomalous zero-curvature (Lax) equation, where the Lax potentials live on the twisted $sl(2)$ Kac-Moody algebra $A_2^{(2)}$. We have used the properties of the Lax potentials under Lorentz transformation and internal  $A_2^{(2)}$ transformations to prove that two-soliton type solution satisfying the  property \rf{phiunderp} under the parity transformation, present an infinity of asymptotically conserved charges. In addition, we have shown that by starting with a two-soliton solution of the integrable Bullough-Dodd model, satisfying  \rf{phiunderp}, one can always construct by a power series expansion on the deformation parameter $\ve$, a solution of the deformed model that also satisfies the parity property \rf{phiunderp} and leading to the asymptotic conservation of the charges. So, the dynamics of the deformed model seems to favor the quasi-integrable sector of the theory.  
 
 The numerical simulations have confirmed the predictions of our analytical calculations. The numerical code was divided in two main parts: the first part uses an exact static one-soliton solution of the integrable Bullough-Dodd model as a seed to constructed a static one-soliton of the deformed model under a relaxation procedure. The second part of the code takes two of such deformed one-solitons far apart and stitches them at the middle point, and then performs the scattering of them, absorbing the radiation at the edges of the grid.   The anomalies of the charge's non-conservation are evaluated to verify their asymptotic conservation. It is interesting to note that the initial configurations of the simulations are built from exact solutions of the Bullough-Dodd model that satisfy the parity  property \rf{phiunderp}. The simulations could well drive them into solutions of the deformed model that do not satisfy  \rf{phiunderp}. However, in all our simulations the real part of the charges were observed to be asymptotically conserved indicating that they do satisfy the property \rf{phiunderp}.  That corroborates the conclusion of our analytical perturbative calculation, mentioned above, that the dynamics of the deformed model favors the solutions satisfying  \rf{phiunderp}.
   
We do not believe that the parity property \rf{phiunderp} is causing the phenomena we call {\em quasi-integrability}, but it seems to be present whenever such non-linear phenomena manifest itself. We need further investigations to fully understand the mechanisms behind the quasi-integrability. That is a quite interesting phenomena that may have several applications in many areas of non-linear sciences.

\vspace{5cm}

{\bf Acknowledgements:} We are grateful to Prof. Wojtek Zakrzewski (Durham Univ.) for many helpful discussions, and to Dr. David Forster for valuable discussions on the early stages of this work. VHA acknowledges the support of a FAPESP scholarship. LAF is partially support by CNPq-Brazil. 

%\vspace{1cm}

\newpage

\appendix

\section{The twisted loop algebra $A_2^{(2)}$}
\label{app:kacmoody}
\setcounter{equation}{0}

We consider here the twisted loop algebra $A_2^{(2)}$, which corresponds to the twisted Kac-Moody algebra $A_2^{(2)}$ with vanishing central term. For our applications it is convenient to work with a realization of such algebra based on $SU(3)$ matrices, and the following basis  
\be
b_{6n+1}\equiv \lambda^n\, b_{1}\qquad 
b_{6n-1}\equiv \lambda^n\, b_{-1}\qquad 
F_{6n+j}\equiv \lambda^n\, F_{j}\qquad j=0,1,2,3,4,5\; ;  \qquad n\in \IZ
\lab{generators}
\ee
where $\lambda$ is the so-called spectral parameter, and 
\br
b_{1}&=&
\left(
\begin{array}{ccc}
 0 & -\frac{1}{\sqrt{3} \omega} & 0 \\
 0 & 0 & -\frac{1}{\sqrt{3} \omega} \\
 \frac{\sqrt{\lambda}}{\sqrt{3} \omega} & 0 & 0
\end{array}
\right)
\qquad \qquad \quad
b_{-1}=
\left(
\begin{array}{ccc}
 0 & 0 & \frac{\omega}{\sqrt{3} \sqrt{\lambda}} \\
 -\frac{\omega}{\sqrt{3}} & 0 & 0 \\
 0 & -\frac{\omega}{\sqrt{3}} & 0
\end{array}
\right)
\nonumber\\
F_{0}&=&
\left(
\begin{array}{ccc}
 1 & 0 & 0 \\
 0 & 0 & 0 \\
 0 & 0 & -1
\end{array}
\right)
\qquad\qquad \qquad\qquad\quad
F_{1}=
\left(
\begin{array}{ccc}
 0 & \frac{1}{\sqrt{3} \omega} & 0 \\
 0 & 0 & \frac{1}{\sqrt{3} \omega} \\
 \frac{2 \sqrt{\lambda}}{\sqrt{3} \omega} & 0 & 0
\end{array}
\right)
\lab{basisa22}\\
F_{2}&=&
\left(
\begin{array}{ccc}
 0 & 0 & 0 \\
 -\frac{\sqrt{\lambda}}{\omega^2} & 0 & 0 \\
 0 & \frac{\sqrt{\lambda}}{\omega^2} & 0
\end{array}
\right)
\qquad\qquad \qquad\qquad
F_{3}=
\left(
\begin{array}{ccc}
 \frac{\sqrt{\lambda}}{\sqrt{3} \omega^3} & 0 & 0 \\
 0 & -\frac{2 \sqrt{\lambda}}{\sqrt{3} \omega^3} & 0 \\
 0 & 0 & \frac{\sqrt{\lambda}}{\sqrt{3} \omega^3}
\end{array}
\right)
\nonumber\\
F_{4}&=&
\left(
\begin{array}{ccc}
 0 & \frac{\sqrt{\lambda}}{\omega^4} & 0 \\
 0 & 0 & -\frac{\sqrt{\lambda}}{\omega^4} \\
 0 & 0 & 0
\end{array}
\right)
\qquad\qquad \qquad\qquad
F_{5}=
\left(
\begin{array}{ccc}
 0 & 0 & \frac{2 \sqrt{\lambda}}{\sqrt{3} \omega^5} \\
 \frac{\lambda}{\sqrt{3} \omega^5} & 0 & 0 \\
 0 & \frac{\lambda}{\sqrt{3} \omega^5} & 0
\end{array}
\right)
\nonumber
\er
where  $\omega$ is such that 
\be
\omega^6=-1\qquad\qquad {\rm i.e.} \qquad\qquad \omega = e^{i\,\(2\,j+1\)\,\pi/6}
\qquad j=0,1,2,3,4,5
\lab{omegadef}
\ee

The commutation relations are given by (independently of the choice of the $\omega$ given in \rf{omegadef})
\be
\sbr{b_{1}}{b_{-1}}=0
\lab{commrel}
\ee
\br
\sbr{b_{1}}{F_k}&=&F_{k+1}\qquad\quad k=0,1,2,3,4\; ; \qquad\qquad \qquad\qquad
\sbr{b_{1}}{F_5}= \lambda \, F_0
\nonumber\\
\sbr{b_{-1}}{F_l}&=& F_{l-1}\qquad\quad\; l=1,2,3,4,5\; ;
\qquad\qquad \qquad \qquad
\sbr{b_{-1}}{F_0}= \lambda^{-1} \, F_5
\nonumber
\er
\br
\sbr{F_0}{F_1}&=&-2\, b_{1}-F_1\qquad \qquad \sbr{F_1}{F_2}=-F_3 
\qquad \qquad\qquad\;\sbr{F_2}{F_4}=-\lambda\, F_0
\nonumber\\
\sbr{F_0}{F_2}&=&-F_2\qquad \qquad\qquad\quad\sbr{F_1}{F_3}=-F_4
\qquad \qquad\qquad\;\sbr{F_2}{F_5}=-2\,\lambda\, b_{1}
\nonumber\\
\sbr{F_0}{F_3}&=&0\qquad \qquad\qquad\qquad\;\sbr{F_1}{F_4}=2\,\lambda\, b_{-1}
\qquad \quad\qquad\;\sbr{F_3}{F_4}=2\,\lambda\, b_{1}-\lambda\,F_1
\nonumber\\
\sbr{F_0}{F_4}&=&F_4\qquad \qquad\qquad\qquad\sbr{F_1}{F_5}=\lambda\,F_0
\qquad \qquad\qquad\;\sbr{F_3}{F_5}=-\lambda\, F_2
\nonumber\\
\sbr{F_0}{F_5}&=&-2\,\lambda\,b_{-1}+F_5\qquad \quad\sbr{F_2}{F_3}=-2\,\lambda\, b_{-1}-F_5
\qquad \;\sbr{F_4}{F_5}=-\lambda\, F_3
\nonumber
\er
The subindices of the generators \rf{generators} correspond to the grades given by the grading operator
\be
d= F_0+6\, \lambda\,\frac{d\;}{d\lambda}\; ; 
\qquad\qquad\qquad\qquad \sbr{d}{{\cal G}_m}=m\, {\cal G}_m
\lab{gradingop}
\ee
with
\br
{\cal G}_{6n}&=&\{\lambda^n\,F_0\}
\qquad\qquad {\cal G}_{6n+1}=\{\lambda^n\,b_1\,,\, \lambda^n F_1\}
\qquad \qquad{\cal G}_{6n+2}=\{ \lambda^n F_2\}
\lab{gradingspaces}\\
{\cal G}_{6n+3}&=&\{\lambda^n\,F_3\}
\qquad \qquad{\cal G}_{6n+4}=\{ \lambda^n F_4\}
\qquad\qquad\qquad {\cal G}_{6n+5}=\{\lambda^{n+1}\,b_{-1}\,,\, \lambda^n F_5\}
\nonumber
\er

\newpage

\begin{figure}[ht]
  \centering
\includegraphics[ width=1\textwidth]{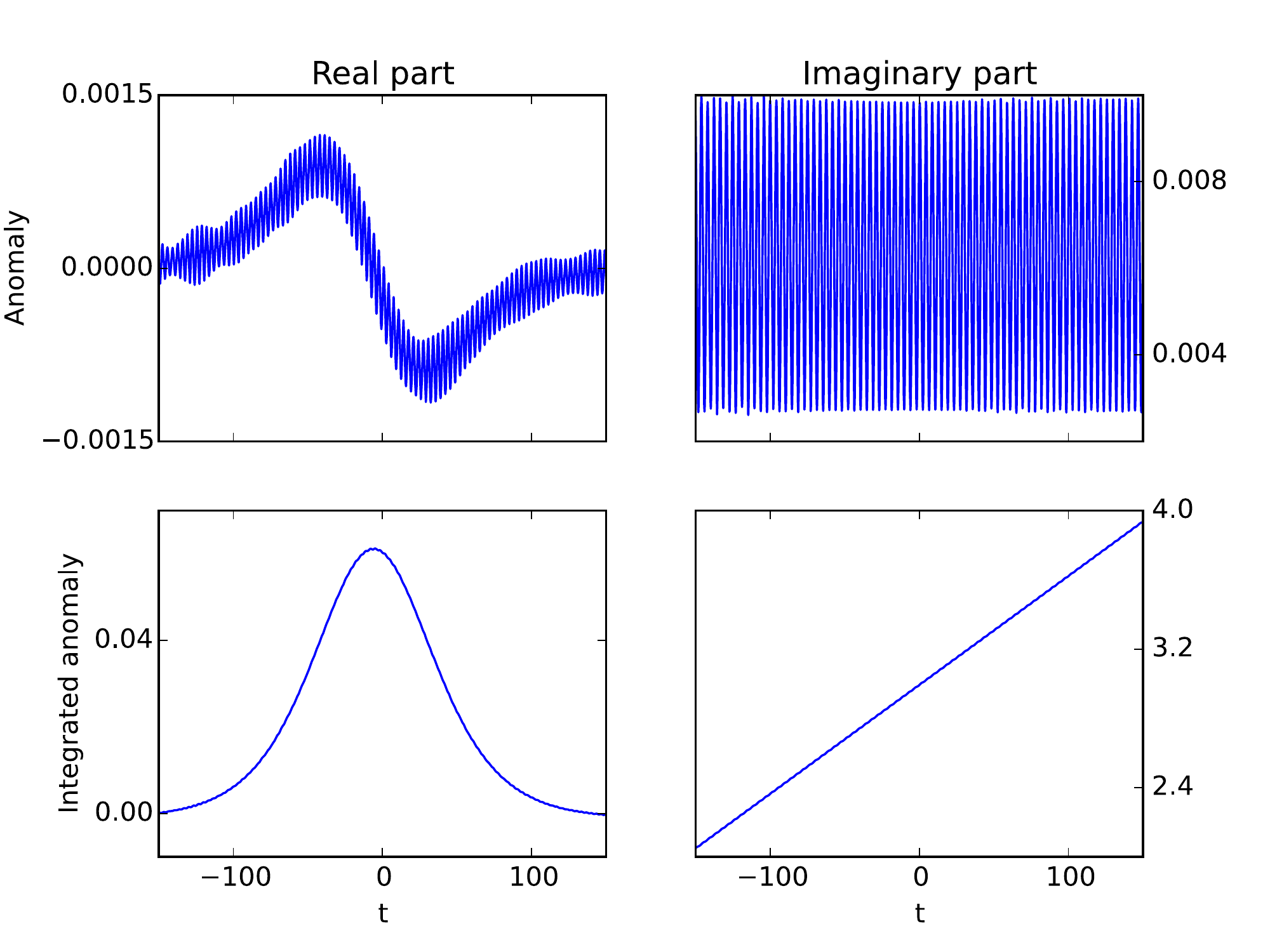}                 
   \caption{Scattering of two kinks for the choice of parameters $\varepsilon=1$, $\xi_1=0.9$ and $\xi_2=0.9$. On the top plots it is shown the real and imaginary parts of the anomaly $\beta^{(5)}$, given in \rf{beta5}, and on the bottom plots are shown the real and imaginary parts of the integrated anomaly $\gamma^{(5)}$, given in \rf{gamma5}. }
  \label{fig:10909}
\end{figure}

\newpage

\begin{figure}[ht]
  \centering
\includegraphics[ width=1\textwidth]{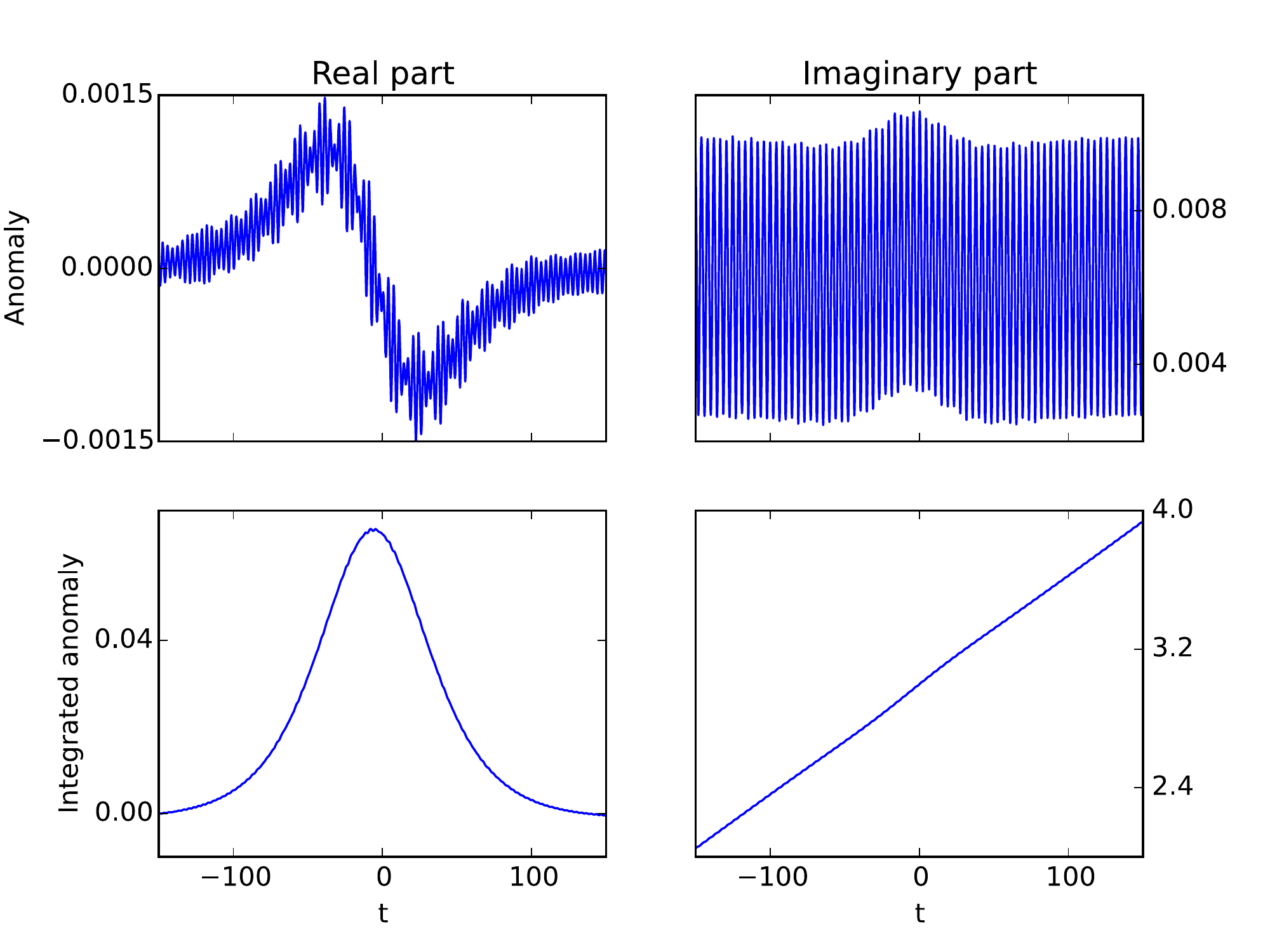}                 
   \caption{Scattering of two kinks for the choice of parameters $\varepsilon=1$, $\xi_1=0.9$ and $\xi_2=1.4$. On the top plots it is shown the real and imaginary parts of the anomaly $\beta^{(5)}$, given in \rf{beta5}, and on the bottom plots are shown the real and imaginary parts of the integrated anomaly $\gamma^{(5)}$, given in \rf{gamma5}.  }
  \label{fig:10914}
\end{figure}
\begin{figure}[ht]
  \centering
\includegraphics[ width=1\textwidth]{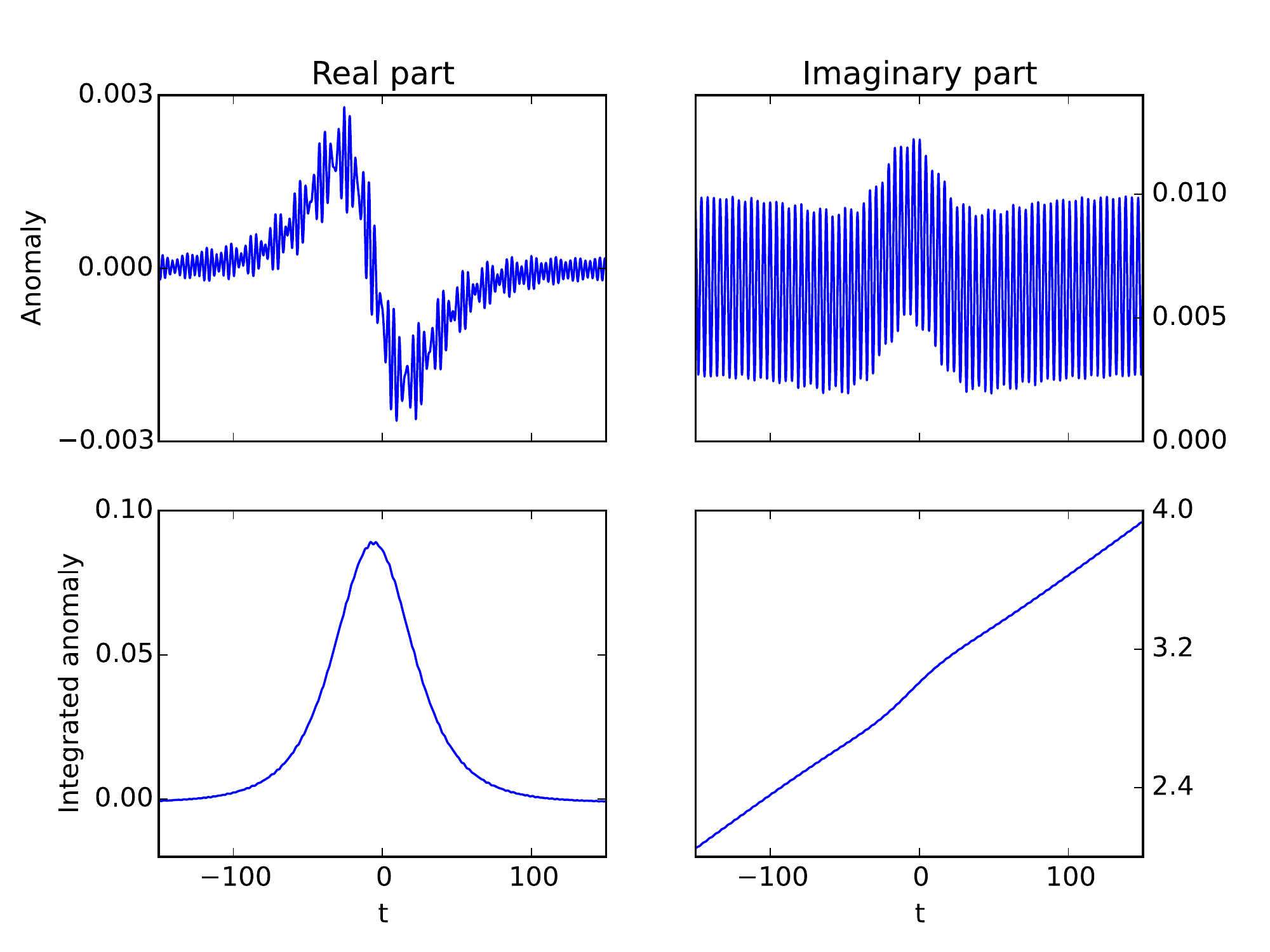}                 
   \caption{Scattering of two kinks for the choice of parameters $\varepsilon=1$, $\xi_1=0.9$ and $\xi_2=2.0$. On the top plots it is shown the real and imaginary parts of the anomaly $\beta^{(5)}$, given in \rf{beta5}, and on the bottom plots are shown the real and imaginary parts of the integrated anomaly $\gamma^{(5)}$, given in \rf{gamma5}. }
  \label{fig:10920}
\end{figure}
\begin{figure}[ht]
  \centering
\includegraphics[ width=1\textwidth]{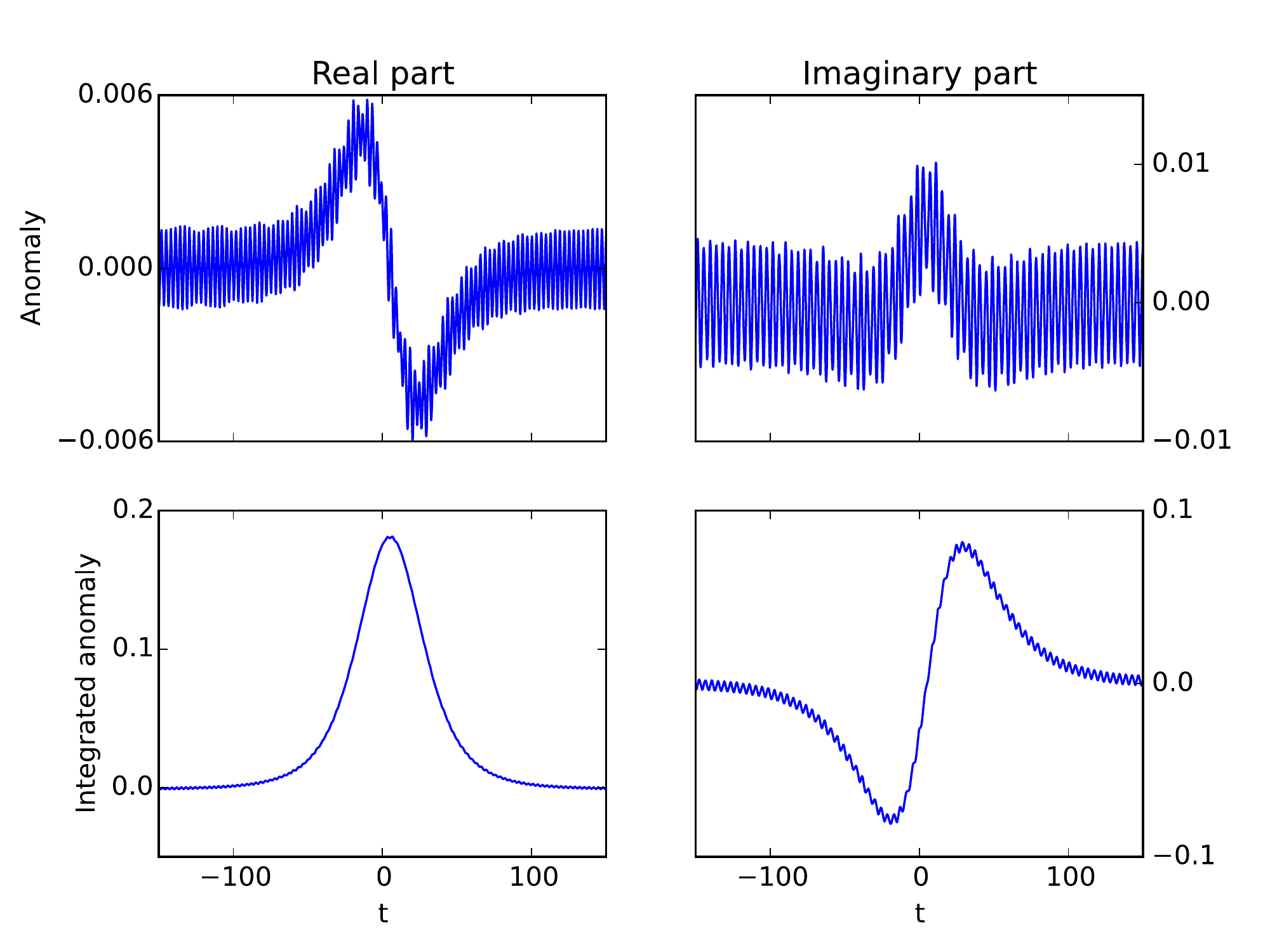}                 
   \caption{Scattering of two kinks for the choice of parameters $\varepsilon=2$, $\xi_1=0.9$ and $\xi_2=2.0$. On the top plots it is shown the real and imaginary parts of the anomaly $\beta^{(5)}$, given in \rf{beta5}, and on the bottom plots are shown the real and imaginary parts of the integrated anomaly $\gamma^{(5)}$, given in \rf{gamma5}. }
  \label{fig:20920}
\end{figure}
\begin{figure}[ht]
  \centering
\includegraphics[ width=1\textwidth]{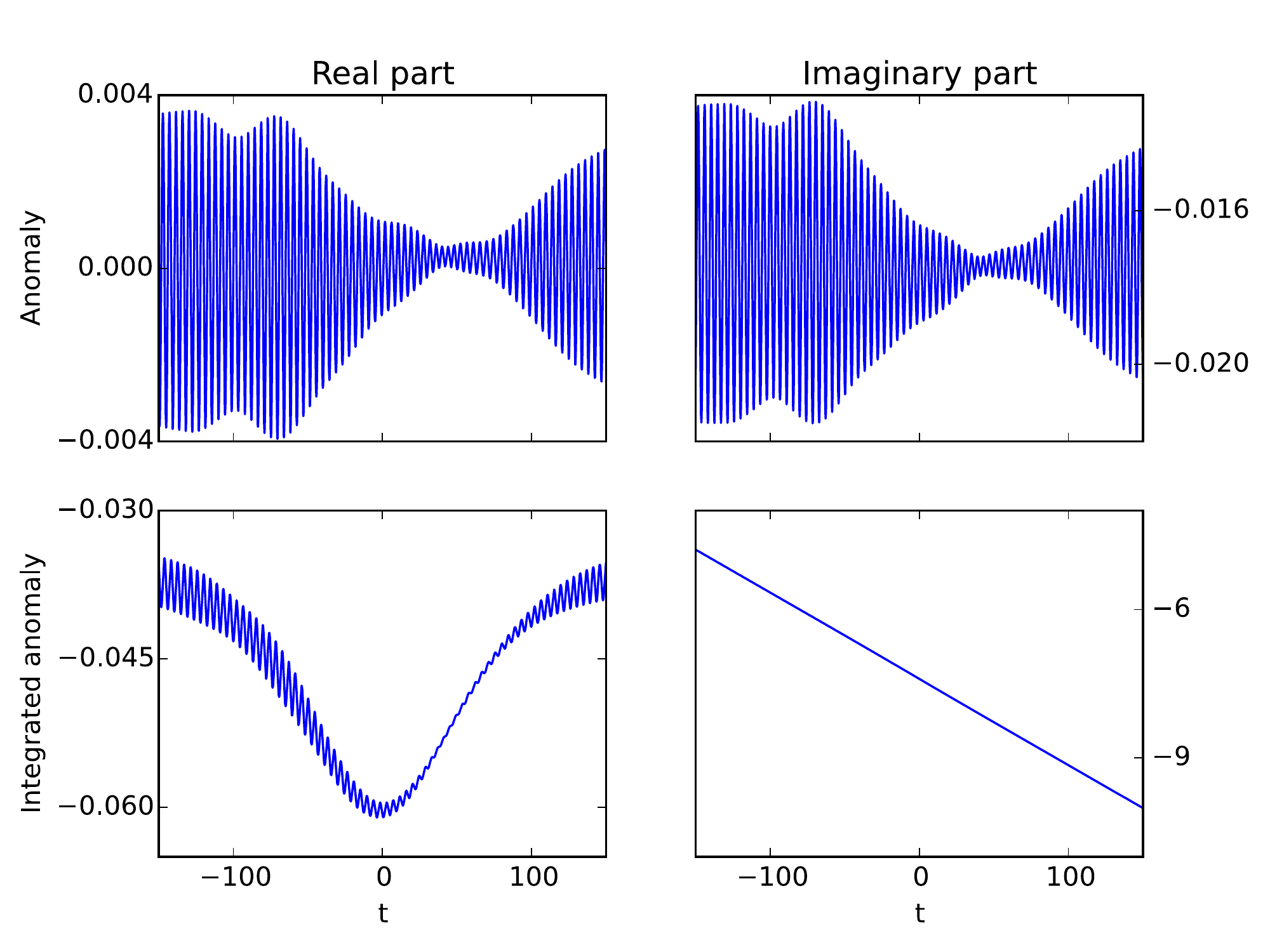}                 
   \caption{Scattering of two kinks for the choice of parameters  $\varepsilon=-1$, $\xi_1=0.9$ and $\xi_2=2.0$. On the top plots it is shown the real and imaginary parts of the anomaly $\beta^{(5)}$, given in \rf{beta5}, and on the bottom plots are shown the real and imaginary parts of the integrated anomaly $\gamma^{(5)}$, given in \rf{gamma5}. }
  \label{fig:m10920}
\end{figure}

\end{document}